\documentclass[aps,prb,twocolumn,showpacs,showkeys,groupedaddress]{revtex4-1}

\usepackage{graphicx}  
\usepackage{dcolumn}   
\usepackage{bm}        
\usepackage{amssymb}   
\usepackage{amsmath}
\usepackage{braket}
\usepackage[mathscr]{euscript}
\usepackage{color}
\usepackage[caption=false]{subfig}
\begin{document}

\title{Study of edge states and conductivity in spin-orbit coupled bilayer graphene}
\author{Priyanka Sinha}
\email{sinhapriyanka2016@iitg.ac.in}
\author{Saurabh Basu}
\email{saurabh@iitg.ac.in}
\affiliation{Department of Physics, Indian Institute of Technology Guwahati\\ Guwahati-781039, Assam, India}
\date{\today}
\begin{abstract}
We present an elaborate and systematic study of the conductance properties of a zigzag bilayer graphene nanoribbon modeled by a Kane-Mele (KM) Hamiltonian. The interplay of the Rashba and the intrinsic spin-orbit couplings with the edge states, electronic band structures, charge and spin transport are explored in details. We have analytically derived the conditions for the edge states for a bilayer KM nanoribbon and show how these modes decay for lattice sites inside the bulk. It is particularly interesting to note that for a finite-size ribbon an even number of zigzag ribbon hosts a finite energy gap at the Dirac points, while the odd ones do not. This asymmetry is present both in presence and absence of a bias voltage that may exist between the layers. The interlayer Rashba spin-orbit coupling, along  with the intralayer intrinsic spin-orbit and intralayer Rashba spin-orbit couplings seem to destroy the quantum spin Hall (QSH) phase where the QSH phase is identified by the presence of a conductance plateau (of magnitude $4e^2/h$) in the vicinity of zero Fermi energy. The plateau is sensitive to the values of the spin-orbit coupling parameters. Further, the spin polarized conductance data reveal that a bilayer KM ribbon is found to be more efficient for spintronic applications compared to a monolayer graphene. Finally, a quick check with experiments is done via computing the effective mass of electrons.
\end{abstract}


\maketitle

\section{\label{sec1}Introduction}
The electronic properties of a single atomic thin layer of graphene\cite{novo,firsov} have attracted much attention in the last few years owing to its unique physical properties, such as, integer quantum Hall effect\cite{zhang,vp}, Klein tunneling\cite{kat}, high carrier mobility\cite{du,bolotin}, half-metallicity\cite{jun,lin} and many more. After the successful fabrication of graphene\cite{novo}, it was observed that bilayer graphene too shows a variety of striking phenomena\cite{sd,novo1}. Many of the properties of a bilayer graphene are analogous to those of a monolayer, such as high thermal conductivity\cite{ghosh,balandin}, high electrical conductivity\cite{dean}, mechanical strength and flexibility\cite{neek,cheng}. However, a bilayer graphene has some additional features that make it distinct from a monolayer graphene. Monolayer graphene has a linear dispersion near the Dirac points\cite{wallace}, whereas in a bilayer graphene, the bands show parabolic dispersion with massive chiral quasiparticles\cite{novo1,cann}. In the absence of an external electric field, the system comprises of four bands, two of them touch each other at zero energy and the other two are separated by an amount equal to the magnitude of the interlayer tunneling, $t_{\perp}$. Moreover, the band gap in bilayer graphene can be opened and tuned by applying a gate voltage externally\cite{neto, castro}.\par
On a parallel front, study of the effects of spin-orbit coupling (SOC) has become one of the most important topics, especially in systems that do not have the surface (or bulk) inversion symmetry. Some of these systems assumably have exciting prospects of spintronic applications\cite{hill, goldman, tombros, wang1} where spin current can be used to transmit dissipationless information. On the other hand, it has been realised that SOC can lead to a new quantum state of matter that supports gapless edge (or surface) states protected by the time-reversal symmetry (TRS), while the bulk remain insulating. It is named as topological insulator (TI) \cite{hasan, qi} or as QSH insulator.\par
There may be different kinds of SOC present in the system due to different physical origins. Mainly, two kinds of SOCs are thought to be relevant in the context of graphene, namely the intrinsic SOC and the Rashba SOC \cite{kane-mele1, kane-mele2}. Kane and Mele showed that the intrinsic SOC favors the QSH phase, whereas the Rashba SOC tends to destroy the QSH phase.\par
From different first-principles studies\cite{min,yao}, the strength of the intrinsic SOC emerges to be of the order of $10^{-3}$ meV. This value is much weaker than the value predicted by Kane-Mele compared to what is needed to realize the topological phase. Nevertheless, owing to its vast and potential applications as spintronic devices, several experimental studies yield the enhanced SOC values which are realized by doping heavy adatoms, such as Indium or Thallium\cite{weeks}. Recently, many other 2D materials have been found with prospects of a tunable SOC, such as silicene, germanene and stanene \cite{lalmi,vogt,chen,tsai,wang} etc. From the first-principles calculation, it is reported that Rashba SOC can be enhanced via doping with 3d or 5d transition-metal atoms \cite{ding,mokrousov}.\par
There are only a very few studies on the effects of spin-orbit coupling in a bilayer graphene so far. Intrinsically, the magnitude of the spin-orbit coupling in a bilayer graphene is about one order of magnitude larger than that in monolayer graphene due to mixing of the $\pi$ and $\sigma$ bands via interlayer hoppings (typically of the order of $0.01-0.1$ meV) \cite{guinea}. A bulk energy gap can be opened by breaking the inversion symmetry via the staggered sublattice potential term and it plays a similar role in a monolayer graphene as that played by the gate bias in a bilayer graphene\cite{niu1, niu2}. It has also been studied that the bias voltage may reduce the bulk energy gap induced by the intrinsic SOC\cite{tao}. The topological phases of a bilayer Kane-Mele model have been studied in details in presence of both SOCs\cite{quio,pan}. The main findings are that a $\mathbb{Z}_{2}$-metallic phase can be achieved with nontrivial $\mathbb{Z}_{2}$ invariant which gives rise to spin helical egde states in presence of the time reversal symmetry, whereas a Chern metallic phase can be achieved with non-trivial Chern invariant. The latter gives rise to chiral egde states with the breaking of time-reversal symmetry by a zeeman-like coupling term \cite{pan}. A stable topological insulator phase can be achieved in gated bilayer graphene in presence of large Rashba SOC\cite{niu}. Further, the study of the band structure reveals that a \textit{mexican hat} feature appears in the vicinity of the Dirac points in presence of SOC and without any bias voltage\cite{smith}.  Moreover, the conventional charge transport in bilayer graphene has been studied earlier \cite{rossi}, but a systematic study of charge and spin transport in a spin-orbit coupled bilayer graphene is still new and hence needs to be explored.\par
It is well-known that graphene nanoribbons (GNRs) can be classified into two kinds: zigzag GNRs (ZGNRs) and armchair GNRs (AGNRs) depending upon the edge termination type\cite{nakada}. It was shown that a monolayer ZGNR supports zero-energy edge states and is dispersionless (flat band) near the Fermi energy\cite{nakada,fujita,sirgist}. Later, it was shown that a bilayer ZGNR also support edge states at zero Fermi energy\cite{Guinea}. \par
In general bilayer graphene can be synthesized in two different configurations, one with carbon ($C$) atoms of the $A$ sublattice in one layer being stacked directly over that of $A$ sublattice of the other layer ($AA$ stacking), or in another, $C$ atoms of $A$ sublattice is stacked over $C$ atoms of $B$ sublattice ($AB$ stacking). In this paper, we explore the roles of different SOCs in a $AB$-stacked bilayer Kane-Mele nanoribbon and emphasize on its various physical properties. We show the effects of spin-orbit coupling on the edge states and the band structure. On the other hand, the transport properties are investigated in order to understand charge conductance and spin polarized conductances, where the latter underscores its spintronic applications.\par
We have organized our paper as follows. In section \ref{hamiltonian}, we present the bilayer Kane-Mele model. In section \ref{surface states}, we derive the fundamental eigenvalue equations that form the backbone of our results for the discussion on the edge states and band structures. In section \ref{bias}, we consider the effect of a bias voltage on the band structure. In section \ref{transport}, we study the quantum transport phenomena which involves charge and spin polarized conductances. We include a connection with the experimental data by computing the effective mass of the electrons in a bilayer graphene and explore how the values are sensitive to the Rashba coupling in section \ref{effective_mass}. Finally, we conclude with a brief summary of our work and its ramifications in section \ref{conclusion}.

\section{Model Hamiltonian}
\label{hamiltonian}
We consider a $AB$-stacked bilayer graphene sheet with zigzag edges which consists of two coupled monolayers of $C$ atoms and hence involves four sublattices labeled by $A_i$ and $B_i$ (where $i= 1$ and $2 $ denote top and bottom layer respectively). The three nearest neighbors vectors in real space are defined by,
$\delta_1 = \big(0, a\big)$;
$\delta_2 = \Big(\frac{\sqrt{3}a}{2}, -\frac{a}{2}\Big)$ and
$\delta_3 = \Big(-\frac{\sqrt{3}a}{2}, -\frac{a}{2}\Big)$,
 $a $ represent the vectors from sublattice sites $A$ to its three nearest sublattice sites $B$. The Kane-Mele (KM) model of a bilayer graphene in presence of a biasing voltage $V$  (neglecting the staggered sublattice potential term) can be written as,
\begin{align}
\mathcal{H} =& \mathcal{H}^T +\mathcal{H}^{B} +\mathcal{H}^{T-B} + V \Bigg( \sum\limits_{i\in{T}, \alpha}c_{i\alpha}^{\dagger} c_{i\alpha}-\sum\limits_{i\in{B},\alpha}c_{i\alpha}^{\dagger} c_{i\alpha} \Bigg) \nonumber \\
=& \mathcal{H}^T +\mathcal{H}^{B} - t_{\perp} \sum\limits_{i\in({T,A}),j\in({B,B}),\alpha}\Big(c_{i\alpha}^{\dagger} c_{j\alpha}+h.c\Big) \nonumber \\
& 
- i \lambda_R^{\perp}\sum\limits_{i\in({T,A}),j\in({B,B}),\alpha\beta}\Bigg(c_{i\alpha}^{\dagger} \left( {\bf s} \times {\bf{\hat{d}}}_{ij}\right)_z c_{j\beta} + h.c\Bigg) \nonumber \\
&
+V \Bigg( \sum\limits_{i\in{T},\alpha}c_{i\alpha}^{\dagger} c_{i\alpha}-\sum\limits_{i\in{B},\alpha}c_{i\alpha}^{\dagger} c_{i\alpha} \Bigg)
\label{eq_1}
\end{align} 
where $\mathcal{H}^T$ and $\mathcal{H}^B$ refer to the Hamiltonians for the top and the bottom graphene layers respectively. $\mathcal{H}^{T-B}$ includes coupling between the top and bottom layers. Since the form of the Hamiltonian depends on the stacking geometry of the layers, we have considered only the hopping between the $A$ site of the top layer and the nearest $B$ site of the bottom layer which is represented by the third term. The subscripts $i$,$j$ label the lattice sites and $\alpha$ denotes the spin index. The interlayer hopping amplitude is denoted by $t_{\perp}$, where $t_{\perp}\simeq 0.4$ eV ($t_{\perp}\ll t$). The fourth term represents the interlayer Rashba coupling arising in presence of a tilted electric field\cite{smith}. The negative sign in the fourth term indicates that the $A$ site of the top layer and nearest $B$ site of the bottom layer are connected by a unit vector -$\bf{\hat{z}}$. The last term is the interlayer bias potential with strength $V$.
The Kane-Mele\cite{kane-mele1,kane-mele2} Hamiltonian contains the following terms for each of the single layers, namely,
\begin{equation}
\mathcal{H}^{T(B)}= \mathcal{H}_{hop}+\mathcal{H}_{ISOC}+ \mathcal{H}_{RSOC}
\end{equation}
 
where,
 \begin{align}
&\mathcal{H}_{hop} = - t\sum\limits_{\langle ij\rangle\alpha}c_{i\alpha}^{\dagger} c_{j\alpha} \nonumber
\\
&
\mathcal{H}_{ISOC}= it_{2}\sum\limits_{\langle \langle ij \rangle\rangle\alpha\beta} \nu_{ij} c_{i\alpha}^{\dagger}s^z_{\alpha\beta} c_{j\beta} \nonumber
\\
&
\mathcal{H}_{RSOC}=  i\lambda_R  \sum\limits_{\langle ij\rangle\alpha\beta}c_{i\alpha}^{\dagger} \left( {\bf s} \times {\bf\hat{d}}_{ij}\right)_z c_{j\beta} \nonumber
\end{align}
The first term $\mathcal{H}_{hop}$ describes the hopping between nearest neighbors with hopping energy $t$ ($t\simeq 2.7$ eV). The second term $\mathcal{H}_{ISOC}$ is the mirror symmetric intrinsic SOC with a coupling strength $t_{2}$. $\nu_{ij}$ = $+1(-1)$ if the electron makes a left (right) move to go from site $j$ to a next neighbor $i$ through their common nearest neighbor. The vector ${\bf\hat{d}}$ points from site $i$ to site $j$ and corresponds to the nearest neighbor vectors. $s^z$ is the $z$-component of Pauli spin matrix. The third term is the nearest neighbor Rashba term which arises due to the perpendicular electric field or interaction with a substrate with a coupling strength $\lambda_R$.

\begin{figure}[h]
\begin{center}
\includegraphics[width=0.45\textwidth]{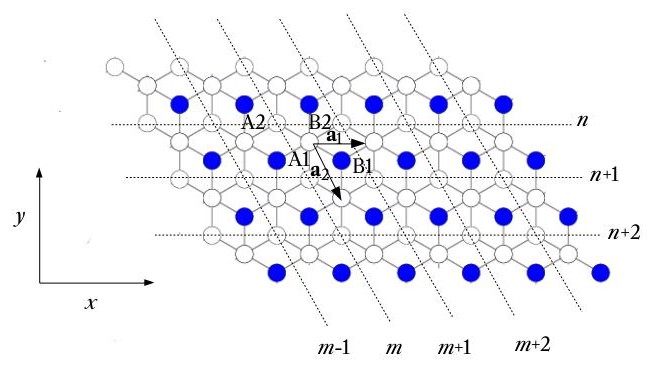}
\caption{(Color online) Bilayer graphene nanoribbon geometry with zigzag edges. The white and blue circles represent the $A$ and $B$ sublattices of the ribbon respectively. $\vec{\bf{a}_1}$ and $\vec{\bf{a}_2}$ are the primitive vectors. ($m, n$) labels the unit cell along $x$ and $y$ direction.}
\label{fig:1}
\end{center}
\end{figure}\vspace{-0.1 cm}


\section{Edge states and band structure}
In the following we show the edge states and the band structure plots for several choices of parameters both in absence and presence of a bias voltage. Initially, we study the model with switching off the biasing term ($V=0$).
\subsection{Zero bias ($V=0$)}
\label{surface states}
In this section, we study the edge state properties of a bilayer graphene. We focus on the bilayer graphene ribbon geometry with zigzag edges where the translational invariance exists along the $x$-axis as shown in Fig.~\ref{fig:1}. The ribbon width is such that it has $N$ unit cells along the $y$-axis (where $n$ $\in 0 $ to $N-1$). We begin with the tight binding model where Eq.~(\ref{eq_1}) can be written (without any of the spin-orbit couplings) in terms of $m$ and $n$ that label the unit cell as shown in Fig.~\ref{fig:1} as\cite{Guinea},
\begin{align} 
\mathcal{H} = &
 - t\sum\limits^{2}_{i=1}\sum\limits_{mn}a^{\dagger}_{i}(m,n)\Big[b_{i}(m,n)+b_{i}(m,n-1) \nonumber
\\
& 
+b_{i}(m-1,n)\Big]+h.c
-t_{\perp}\sum\limits_{mn}a^{\dagger}_{1}(m,n) b_{2}(m,n) +h.c 
\label{h1}
\end{align} 
We use periodic boundary condition along the $x$-direction. Using the momentum representation of the electron operators and solving the time-independent Schr$\ddot{o}$dinger equation, we get the following four eigenvalue equations corresponding to both $A$ and $B$ sublattices, 
\begin{align}
E \alpha_1\big(k_{x},n\big) &=  -t\Big[D_k \beta_1(k_x,n)+\beta_1(k_x, n-1)\Big]\nonumber
\\
& 
-t_{\perp} \beta_2(k_x,n) \nonumber
\\
 E \alpha_2\big(k_{x},n\big) & = -t\Big[D_k\beta_2(k_x,n)+\beta_2(k_x, n-1)\Big] \nonumber
\\
E \beta_1(k_{x},n\big) &= -t\Big[D_k\alpha_1(k_x,n)+\alpha_1(k_x, n+1)\Big] \nonumber
\\
E \beta_2(k_{x},n\big) &=  -t\Big[D_k\alpha_2(k_x,n)+\alpha_2(k_x, n+1)\Big] \nonumber
\\
&
- t_{\perp} \alpha_1(k_x,n)
\end{align}
where $\alpha_i$, $\beta_i$ refer to the amplitudes corresponding to the $A$ and $B$ sublattices and $D_k=2  \cos\Big(\frac{\sqrt{3}k_x}{2}\Big) $. We have chosen the basis as,
\begin{align}
|\psi_{k}\rangle=&\sum\limits_{n, \sigma}\sum\limits^{2}_{i=1}\Big[\alpha_i(k, n, \sigma)|a_i, k, n, \sigma\rangle \nonumber
\\
&
+\beta_i(k, n, \sigma)|b_i, k, n, \sigma\rangle \Big].
\end{align}
In case of a bilayer nanoribbon, the boundary condition is,
\begin{equation}
\alpha_1(k_x,N)= \alpha_2(k_x,N)=\beta_1(k_x,-1)=\beta_2(k_x,-1)=0
\label{b.c1}
\end{equation}
To make $k_x$ a dimensionless quantity, we have absorbed the lattice spacing $a$ into the definition of $k_x$. Using the above boundary condition (Eq.~(\ref{b.c1})) and applying the induction method, we finally obtain the following matrix equations for the amplitudes of the wavefunction at $A$ and $B$ sublattices as,

\begin{gather}
 \begin{bmatrix} \alpha_1(k_x,n) \\ \alpha_2(k_x,n) \end{bmatrix}
 =
  \begin{bmatrix}
   D^{n}_{k} & 0 \\
   - n D^{n-1}_{k} \frac{t_\perp}{t} & D^{n}_{k} 
   \end{bmatrix} \begin{bmatrix} \alpha_1(k_x,0) \\ \alpha_2(k_x,0) \end{bmatrix} 
\end{gather}

\begin{gather}
 \begin{bmatrix} \beta_2(k_x,N-n-1) \\ \beta_1(k_x,N-n-1) \end{bmatrix}
 =
  \begin{bmatrix}
   D^{n}_{k} & 0 \\
   - n D^{n-1}_{k} \frac{t_\perp}{t} & D^{n}_{k} 
   \end{bmatrix} \begin{bmatrix} \beta_2(k_x,N-1) \\ \beta_1(k_x,N-1) \end{bmatrix} 
\end{gather}
We choose two linearly independent initial vectors $ [\alpha_1(k_x,0) , 0]$ and $ [0 , \alpha_2(k_x,0)]$ to compute the edge states. Using the Gram-Schmidt orthogonalization process (taking $N\rightarrow \infty$), we finally obtain for the amplitudes at A as, 
\begin{align}
&\alpha_1(k_x, n)= 0  \nonumber  \\
&\alpha_2(k_x, n)= D^{n}_{k} \alpha_2(k_x,0)
\label{edge_1}
 \end{align}    
and
\begin{align}
&\alpha_1(k_x, n)= \alpha_1(k_x,0) D^{n}_{k} \nonumber  \\
&\alpha_2(k_x, n)= -\alpha_1(k_x,0) D^{n-1}_{k} \frac{t_{\perp}}{t}\left(n - \frac{ D^{2}_{k}}{1 - D^{2}_{k}}\right)
\label{edge_2}
\end{align}
and for B sublattice,
\begin{align}
&\beta_2(k_x, N-n-1)= 0  \nonumber  \\
&\beta_1(k_x, N-n-1)= D^{n}_{k} \beta_1(k_x,N-1)
\label{edge_3}
 \end{align}    
and
\begin{align}
\beta_2(k_x, N-n-1)=& \beta_2(k_x,N-1) D^{n}_{k} \nonumber  \\
\beta_1(k_x, N-n-1)=&-\beta_2(k_x,N-1) D^{n-1}_{k} \frac{t_{\perp}}{t} \nonumber
 \\
& \left(n - \frac{ D^{2}_{k}}{1 - D^{2}_{k}}\right)
\label{edge_4}
\end{align}
which represent the orthonormalized zero-energy edge states for a bilayer graphene. From Eq.~(\ref{edge_1}), we can see that the amplitudes corresponding to the sites belonging to the A sublattices of layer 1 vanish and that of layer 2 are finite. This implies that the corresponding equations and hence their solutions are applicable for a monolayer graphene, whereas Eq.~(\ref{edge_2}) refers to the respective solutions for a bilayer graphene. Here the sites for both the layers are found to have non-vanishing amplitudes and that the amplitudes in layer 2 are connected to those in layer 1 via $t_{\perp}$. Similarly, the same applies for the B sublattice through Eq.~(\ref{edge_3}) and (\ref{edge_4}) (in case if we increase the sheet from the other side of the ribbon). \par
Now we additionally consider the intralayer intrinsic spin-orbit interaction superposed on our tight binding model. Since intrinsic SOC involves next to nearest neighbor coupling in a plane, it does not affect the interlayer Hamiltonian. Hence using the same basis, we get eight eigenvalue equations corresponding to the spin up and spin down states for $A_i$ and $B_i$ ($i=1, 2$). They are,
\begin{widetext}
\begin{align}
 E\alpha_{1\pm}\big(k_{x},n\big)
 =&
 - t\Big[D_k\beta_{1\pm}(k_x,n)
 + \beta_{1\pm}(k_x, n-1)\Big]-t_{\perp} \beta_{2\pm}(k_x,n) 
 \mp 2 t_2\Big[P_k\alpha_{1\pm}(k_x,n)-M_k\big\{\alpha_{1\pm}(k_x,n-1) \nonumber
 \\
 & 
 +\alpha_{1\pm}(k_x,n+1)\big\}\Big] \nonumber
 \\ 
 E\alpha_{2\pm}\big(k_{x},n\big)
 =&
 - t\Big[D_k\beta_{2\pm}(k_x,n)
 + \beta_{2\pm}(k_x, n-1)\Big]
\mp 2 t_2\Big[P_k\alpha_{2\pm}(k_x,n)-M_k\big\{\alpha_{2\pm}(k_x,n-1) 
+\alpha_{2\pm}(k_x,n+1)\big\}\Big] \nonumber
 \\
 E\beta_{1\pm}\big(k_{x},n\big)
 =&
 -t\Big[D_k\alpha_{1\pm}(k_x,n)
 + \alpha_{1\pm}(k_x, n+1)\Big]
\pm 2  t_2\Big[P_k\beta_{1\pm}(k_x,n)-M_k\big\{\beta_{1\pm}(k_x,n-1)+\beta_{1\pm}(k_x,n+1)\big\}\Big] \nonumber
 \\
 E\beta_{2\pm}\big(k_{x},n\big)
 =&
 -t\Big[D_k\alpha_{2\pm}(k_x,n)
 + \alpha_{2\pm}(k_x, n+1)\Big] - t_{\perp} \alpha_{1\pm}(k_x,n)
\pm 2  t_2\Big[P_k\beta_{2\pm}(k_x,n)-M_k\big\{\beta_{2\pm}(k_x,n-1) \nonumber
\\
&
+\beta_{2\pm}(k_x,n+1)\big\}\Big]
\label{eq.2} 
\end{align}
\end{widetext}
where the '$\pm$' sign in the subscript denote the up and down spins for both sublattices in each layer and 
$
D_k =  2 \cos\Big(\frac{\sqrt{3}k_x}{2}\Big);  \nonumber 
P_k = \sin\left(\sqrt{3}k_x\right);  \nonumber 
M_k = \sin\Big(\frac{\sqrt{3}k_x}{2}\Big). \nonumber
$ The total $z$-component of the spin remains conserved here.\par
Next we consider the Rashba SOC which leads to spin mixing in presence of a tilted electric field. The intralayer and the interlayer Rashba couplings can now be included and hence the amplitudes, $\alpha_{i_{\pm}}$ and $\beta_{i_{\pm}}$ are obtained from the following sets of equations.
\begin{widetext} 
\begin{align}
 E\alpha_{1\pm}\big(k_{x},n\big)
 =&
 - t\Big[D_k\beta_{1\pm}(k_x,n)
 + \beta_{1\pm}(k_x, n-1)\Big]-t_{\perp} \beta_{2\pm}(k_x,n) 
 \mp 2 t_2\Big[P_k\alpha_{1\pm}(k_x,n) -M_k\big\{\alpha_{1\pm}(k_x,n-1) \nonumber
\\
& 
+\alpha_{1\pm}(k_x,n+1)\big\}\Big] +i\lambda_R\Big[N_{\pm}\beta_{1\mp}(k_x,
  n)-\beta_{1\mp}(k_x,n-1)\Big]\mp\lambda^{\perp}_{R}\beta_{2\mp}(k_x,n)   \nonumber 
 \\  
 E\alpha_{2\pm}\big(k_{x},n\big)
 =&
 - t\Big[D_k\beta_{2\pm}(k_x,n)
 + \beta_{2\pm}(k_x, n-1)\Big] 
  \mp 2 t_2\Big[P_k\alpha_{2\pm}(k_x,n)-M_k\big\{\alpha_{2\pm}(k_x,n-1)+\alpha_{2\pm}(k_x,n+1)\big\}\Big] \nonumber
\\
& 
+i\lambda_R\Big[N_{\pm}\beta_{2\mp}(k_x,
  n)-\beta_{2\mp}(k_x,n-1)\Big] \nonumber 
 \\
 E\beta_{1\pm}\big(k_{x},n\big)
 =&
 -t\Big[D_k\alpha_{1\pm}(k_x,n)
 + \alpha_{1\pm}(k_x, n+1)\Big]  
 \pm 2  t_2\Big[P_k\beta_{1\pm}(k_x,n)-M_k\big\{\beta_{1\pm}(k_x,n-1)+\beta_{1\pm}(k_x,n+1)\big\}\Big] \nonumber
 \\
 &
-i\lambda_R\Big[N_{\mp}\alpha_{1\mp}(k_x,
  n)-\alpha_{1\mp}(k_x,n+1)\Big] \nonumber
  \\
 E\beta_{2\pm}\big(k_{x},n\big)
 =&
 -t\Big[D_k\alpha_{2\pm}(k_x,n)
 + \alpha_{2\pm}(k_x, n+1)\Big] - t_{\perp} \alpha_{1\pm}(k_x,n)
 \pm 2  t_2\Big[P_k\beta_{2\pm}(k_x,n)-M_k\big\{\beta_{2\pm}(k_x,n-1) \nonumber
 \\
 &
+\beta_{2\pm}(k_x,n+1)\big\}\Big]-i\lambda_R\Big[N_{\mp}\alpha_{2\mp}(k_x,
  n)-\alpha_{2\mp}(k_x,n+1)\Big]\pm\lambda^{\perp}_{R}\alpha_{1\mp}(k_x,n)
\label{eq.3} 
\end{align}
\end{widetext}
where $N_{\pm}= \left[\cos(\frac{\sqrt3k_x}{2})\pm \sqrt3 \sin(\frac{\sqrt3 k_x}{2})\right]$. We have solved these equations (namely, Eq.~(\ref{eq.2}) and Eq.~(\ref{eq.3})) numerically using the boundary condition in Eq.~(\ref{b.c1}) for our purpose. These equations become particularly simple corresponding to $k_x= \frac{\pi}{\sqrt3}$ where $D_k$ and $P_k$ (see their definition above) vanish.\par 
Next we discuss the results obtained via solving the eigenvalue equations, namely, Eq.~(\ref{edge_2}), (\ref{edge_4}), (\ref{eq.2}) and (\ref{eq.3}). All the energies are measured in units of inplane hopping $t$. All the parameters in our paper are somewhat overestimated. The reason being that we are interested to see observable effects, where the experimental (or the first principles) values will yield effects that may not be well pronounced. We start with the results for the tight binding case (Eq.~(\ref{edge_2}) and Eq.~(\ref{edge_4})). We have fixed our interlayer tight binding coupling parameter, $t_{\perp}=0.2$ and have considered different parameter values for the intralayer intrinsic SOC, $t_2$ and the inplane and interlayer Rashba couplings, namely $\lambda_R$ and $\lambda^{\perp}_R$ respectively. For example, the actual value of $t_\perp$ is around $0.4$ eV which is approx $0.14t$ $(t=2.7$ eV). Nevertheless, the SOC parameters considered by us are indeed higher (usually one or two order) compared to the actual values. But then as we said earlier these SOC values can be enhanced by using heavily adatoms, we proceed with the overestimated values without trepidation.\par
To study the surface (or edge) state properties for different sublattices, we have plotted the charge densities as a function of $n$ (n being the site index) for different cases. We have also computed the electronic energy dispersion to provide support to the corresponding results for the edge states for a few orbitals, that is, corresponding to small values of $N$. \par

\begin{figure}[!ht]
 \centering
   \subfloat[]{\includegraphics[trim=0 0 0 0,clip,width=0.24\textwidth]{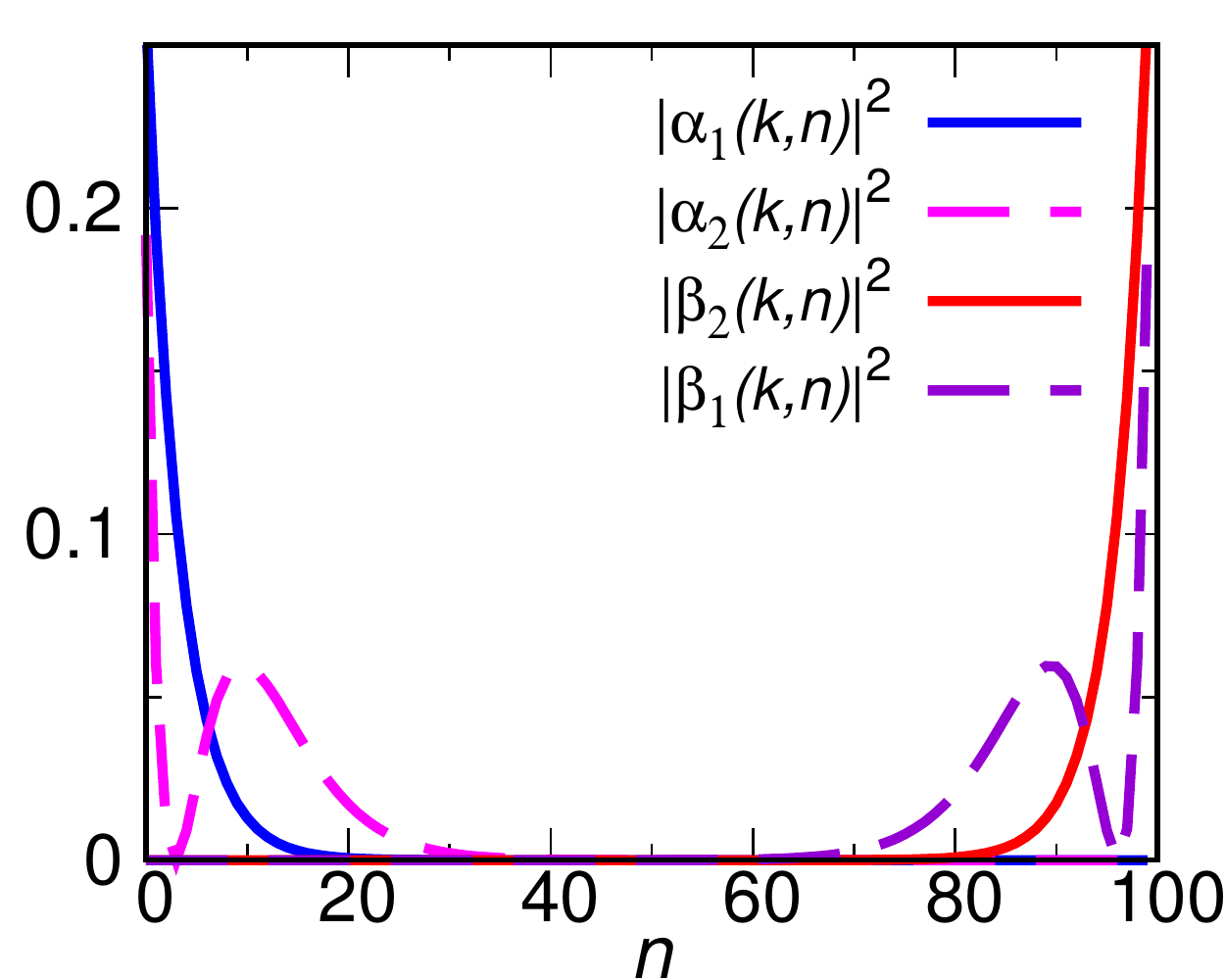}\label{fig:2a}} 
   \subfloat[]{\includegraphics[trim=0 0 0 0,clip, width=0.245\textwidth]{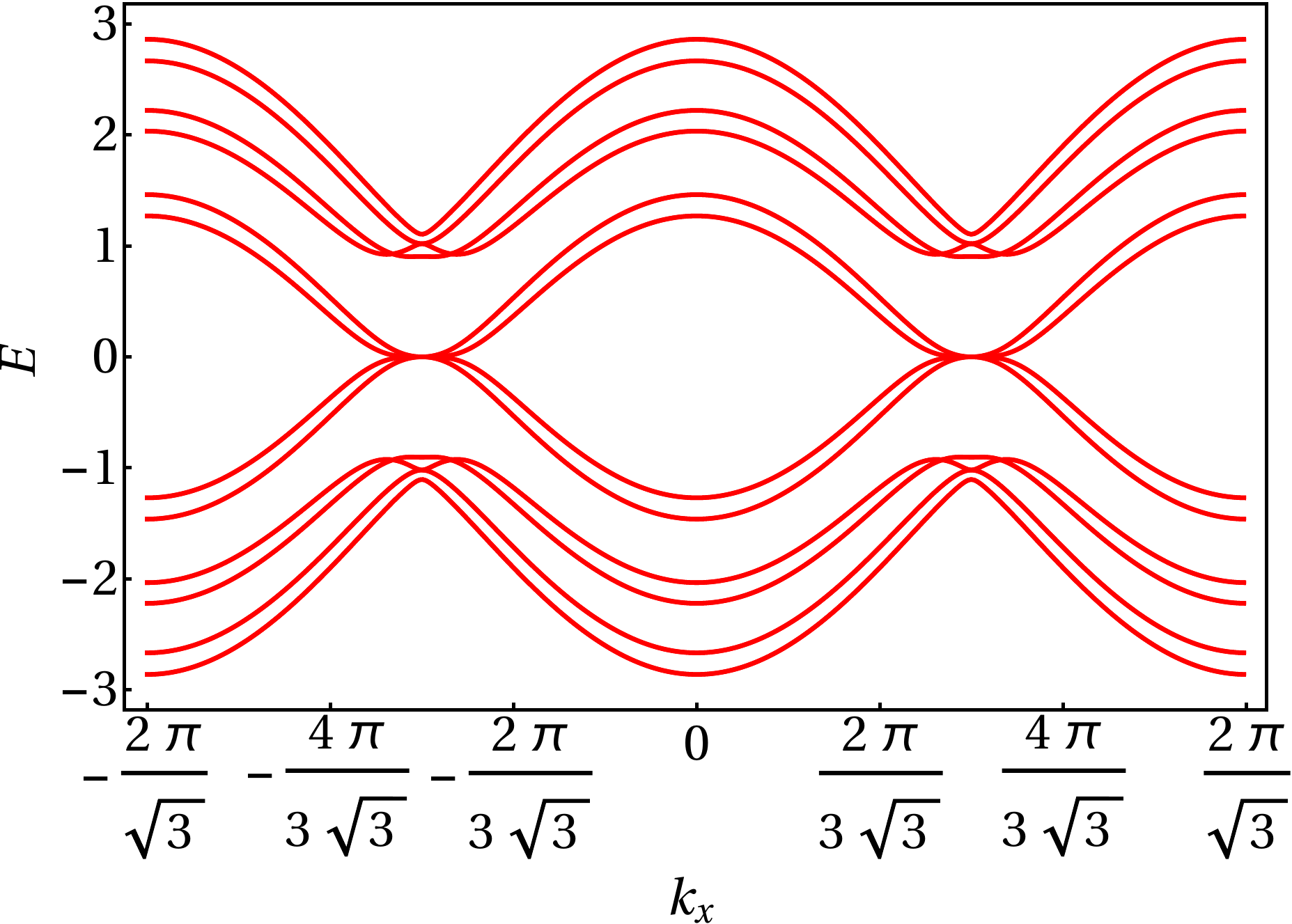}\label{fig:2b}}  
  \caption{(color online) (a) Charge density as a function of site index, $n$ at $k_x= \frac{2.15 \pi}{3\sqrt3}$ and (b) the band structure for $N=3$. Here, we set $t_\perp$ = 0.2.}
\label{fig:2}
\end{figure}
In Fig.~\ref{fig:2a}, we plot Eq.~(\ref{edge_2}) and Eq.~(\ref{edge_4}) for the top and bottom layers which yield that the edge states exist at the zigzag edges at zero energy. They are strictly localized at the edges. The sublattices $\alpha_1$ and $\beta_2$ fall off gradually at both the edges of the ribbon, whereas $\alpha_2$ and $\beta_1$ show different nature than the other two, namely, $\alpha_1$ and $\beta_2$. It can also be seen that the penetration depth of the amplitudes into the bulk get enhanced because of the linear dependence of $\alpha_2$ on $n$.
\begin{figure}[!ht]
 \centering
   \subfloat{\includegraphics[trim=0 0 0 0,clip, width=0.28\textwidth]{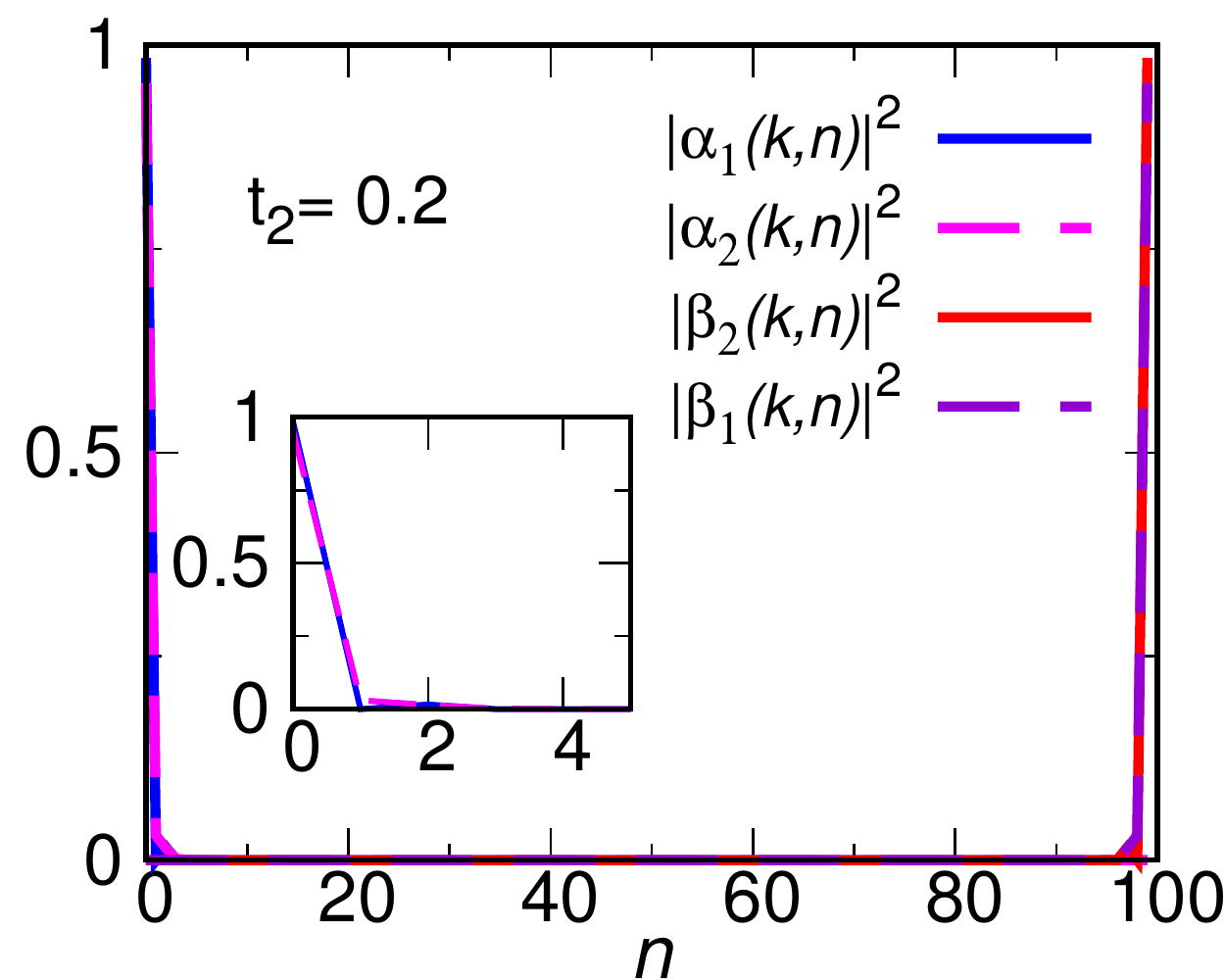}\label{fig:3}}  
  \caption{(color online) Charge density as a function of site index, $n$ at $k_x= \frac{ \pi}{\sqrt3}$. Here, we set $t_\perp = 0.2$ and $t_2= 0.2$.}
\label{fig:3}
\end{figure}

In Fig.~\ref{fig:2b}, we have plotted the band structure for $N=3$ by solving the tight-binding Hamiltonian. Although the flat band is observed at $E=0$ in the momentum intervals  
$\frac{2\pi}{3\sqrt3}\leqslant k_x \leqslant \frac{4\pi}{3\sqrt3} $ and $-\frac{2\pi}{3\sqrt3}\leqslant k_x \leqslant -\frac{4\pi}{3\sqrt3} $, in contrast to a monolayer\cite{priyanka}, there are four flat bands in the above mentioned $k_x$ range which correspond to four edge states for the bilayer and they remain dispersionless. The particle-hole symmetry is conserved and the energy spectrum remains spin-degenerate. \\

\begin{figure}[!ht]
 \centering
   \subfloat[]{\includegraphics[trim=0 0 0 0,clip,width=0.24\textwidth]{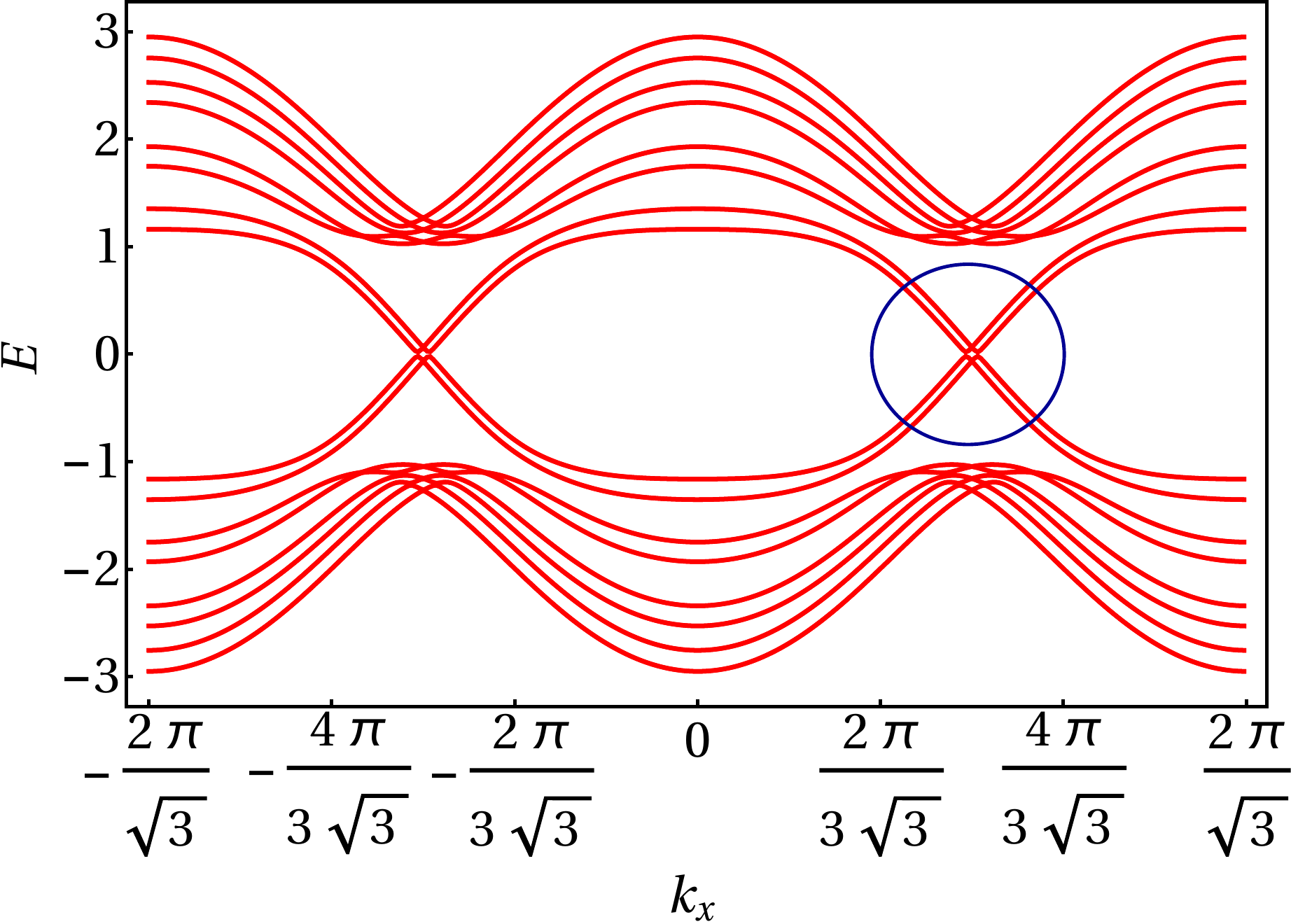}\label{fig:4a}}\hspace*{0.1 cm}
   \subfloat[]{\includegraphics[trim=0 -30 0 0,clip, width=0.16\textwidth]{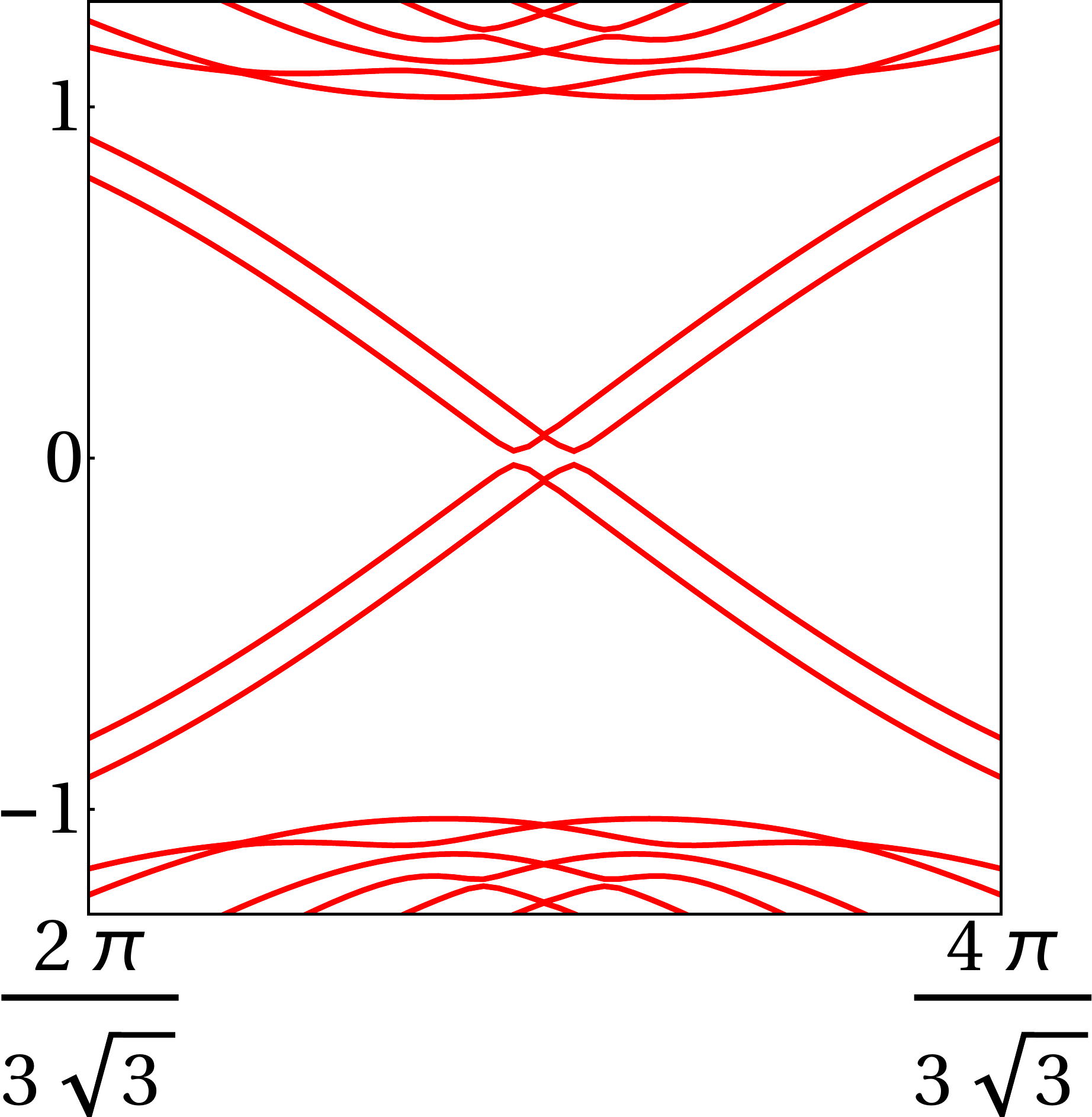}\label{fig:4b}}\\
    \subfloat[]{\includegraphics[trim=0 -30 0 0,clip, width=0.34\textwidth]{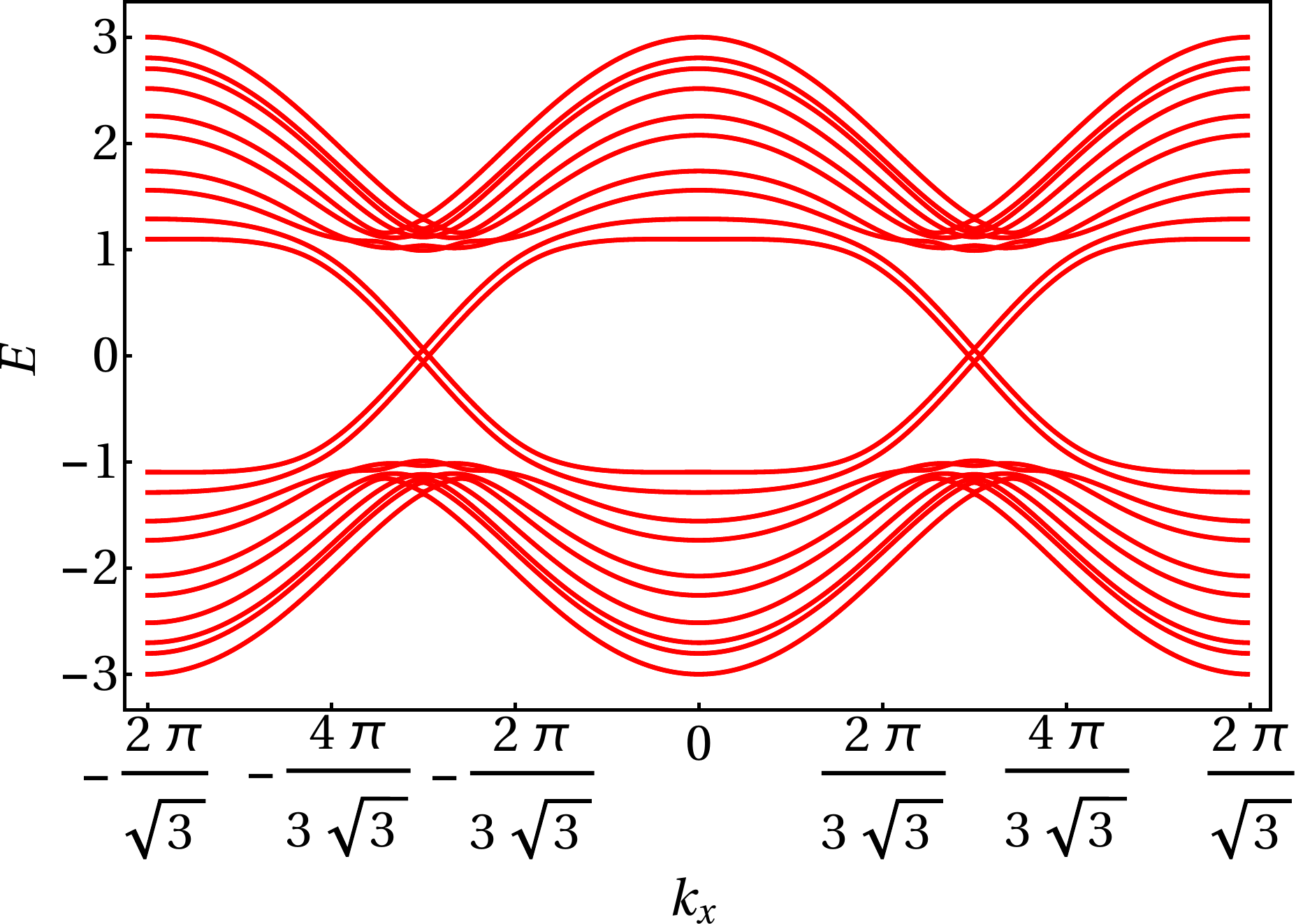}\label{fig:4c}}

  \caption{(color online) The band structure for different values of $N$ (a) $N=4$, (b) zoomed view of the blue circle as shown in (a) and (c) $N=5$ (here there is no gap). Here, we set $t_\perp= 0.2$ and $t_2 = 0.2$.}
\label{fig:4}
\end{figure} 
Next we have considered only the intralayer intrinsic SOC as given in Eq.~(\ref{eq_1}). Fig.~\ref{fig:3} shows the charge density plot at the Dirac point for intrinsic SOC strength $t_2= 0.2$ at an energy close to zero. As compared to a pristine bilayer, the amplitudes for the A sublattice in layer 1 and the B sublattice in layer 2 fall off sharply at the two opposite edges of the ribbon, while the amplitudes for the A sublattice in layer 2 and the B sublattice in layer 1 fall off more gradually as is shown in the inset plot. However, these edge states are topologically protected by the TRS.\par
Fig.~\ref{fig:4} shows the band stucture for different values of $N$, namely an even $N$ ($N=4$) and an odd $N$ ($N=5$) for $t_2=0.2$. It can be seen that there is an odd-even asymmetry in presence of the intrinsic SOC. Fig.~\ref{fig:4a} shows that there is an opening of the band gap of small magnitude at the two Dirac points for $N=4$. It is well-known that the backscattering is forbidden between the time-reversed pairs in a QSH state. Due to the finite-size effects, backscattering still may be possible, which demonstrate that though the time-reversal symmetry is trying to keep gapless edge state, a small gap may be open up. However, for $N=5$ (Fig.~\ref{fig:4c}), one sees the closing of the gap for the same value of the intrinsic coupling constant. The above odd-even scenario depends on the ribbon width. For large values of $N$, such descrepancies will cease to exist. This asymmetry occurs only for a ZGNR may be attributed due to its configuration in the sense that the even ZGNRs ($N$ = even) are in "zigzag/zigzag" configuration, while the odd ZGNRs ($N$ = odd) are in "zigzag/antizigzag" configuration\cite{Li} where the gap in the former case appears owing to a lack of sublattice translational invariance which renders an effective interedge tunneling.

\begin{figure}[!ht]
 \centering
   \subfloat[]{\includegraphics[trim=0 0 0 0,clip,width=0.255\textwidth]{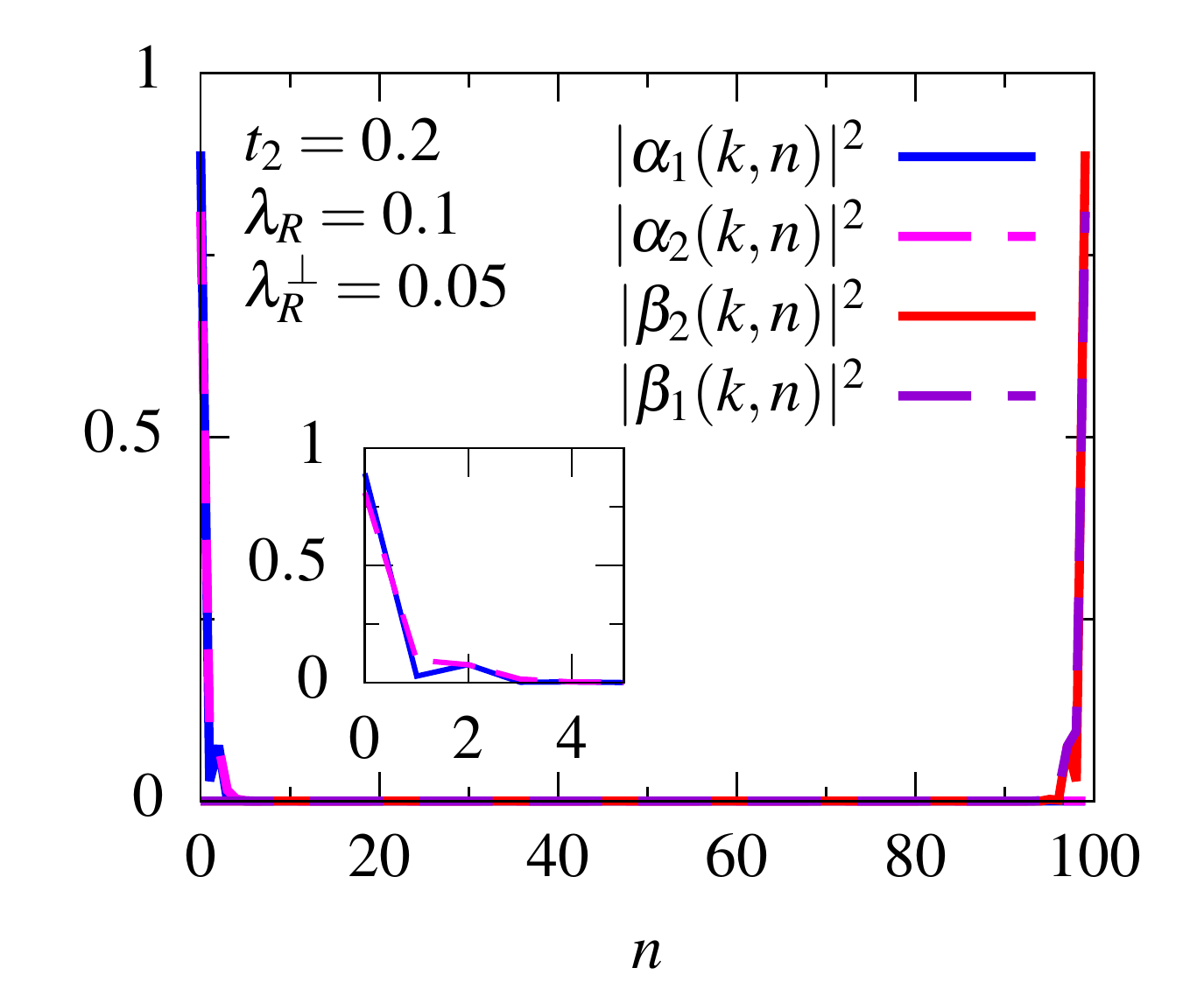}\label{fig:5a}} \hspace*{-0.4 cm}
   \subfloat[]{\includegraphics[trim=0 0 0 0,clip, width=0.255\textwidth]{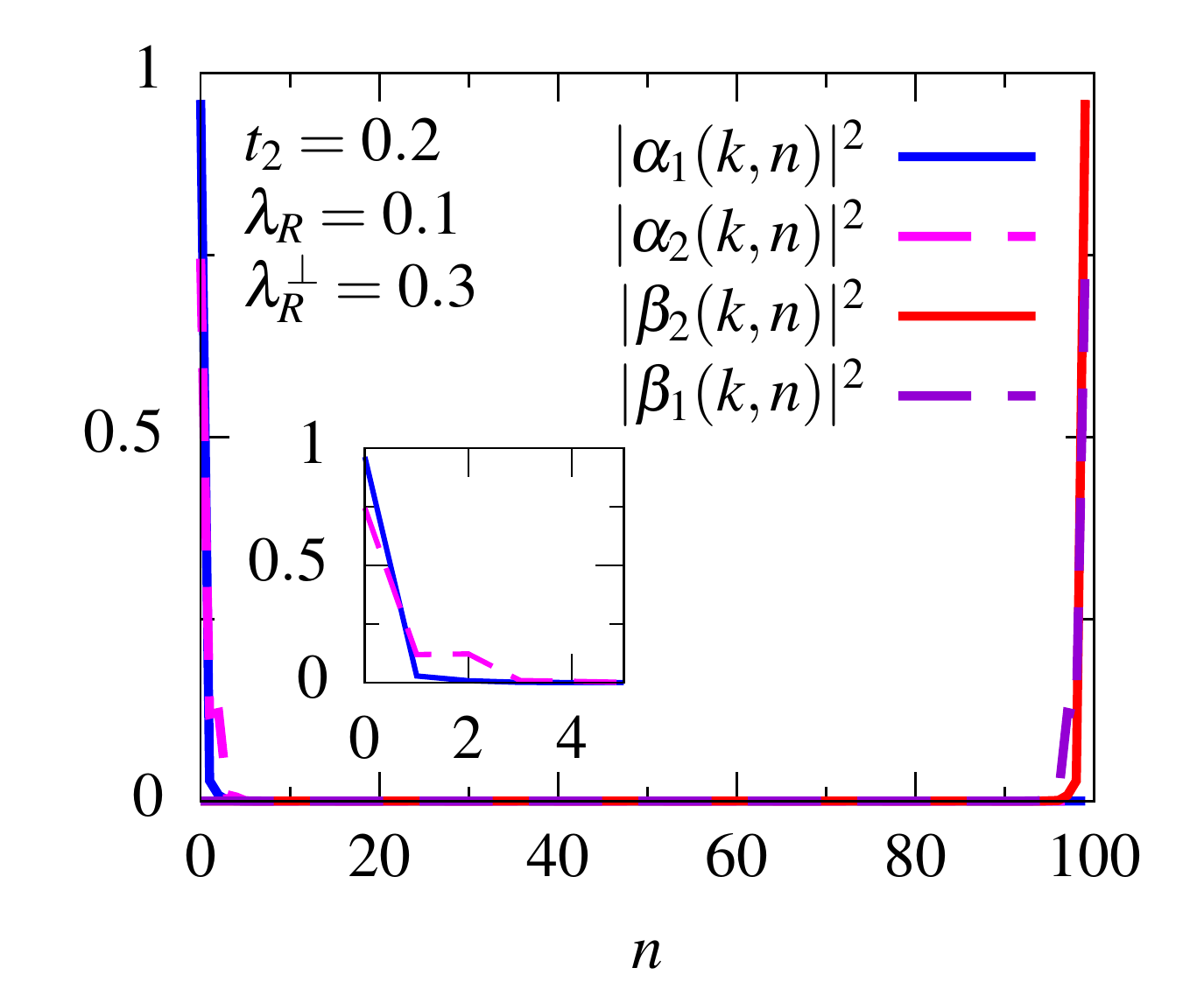}\label{fig:5b}}  
  \caption{(color online) Charge density as a function of site index, $n$ at $k_x=\frac{\pi}{\sqrt3}$ for different values of $\lambda_R^{\perp}$ (a) $\lambda^{\perp}_{R}=0.05$  (b) $\lambda^{\perp}_{R}=0.3$. Here, we set $t_\perp=0.2$, $t_2= 0.2$ and $\lambda_R=0.1$.}
\label{fig:5}
\end{figure}

\begin{figure}[!ht]
 \centering
   \subfloat[]{\includegraphics[trim=0 0 0 0,clip,width=0.24\textwidth]{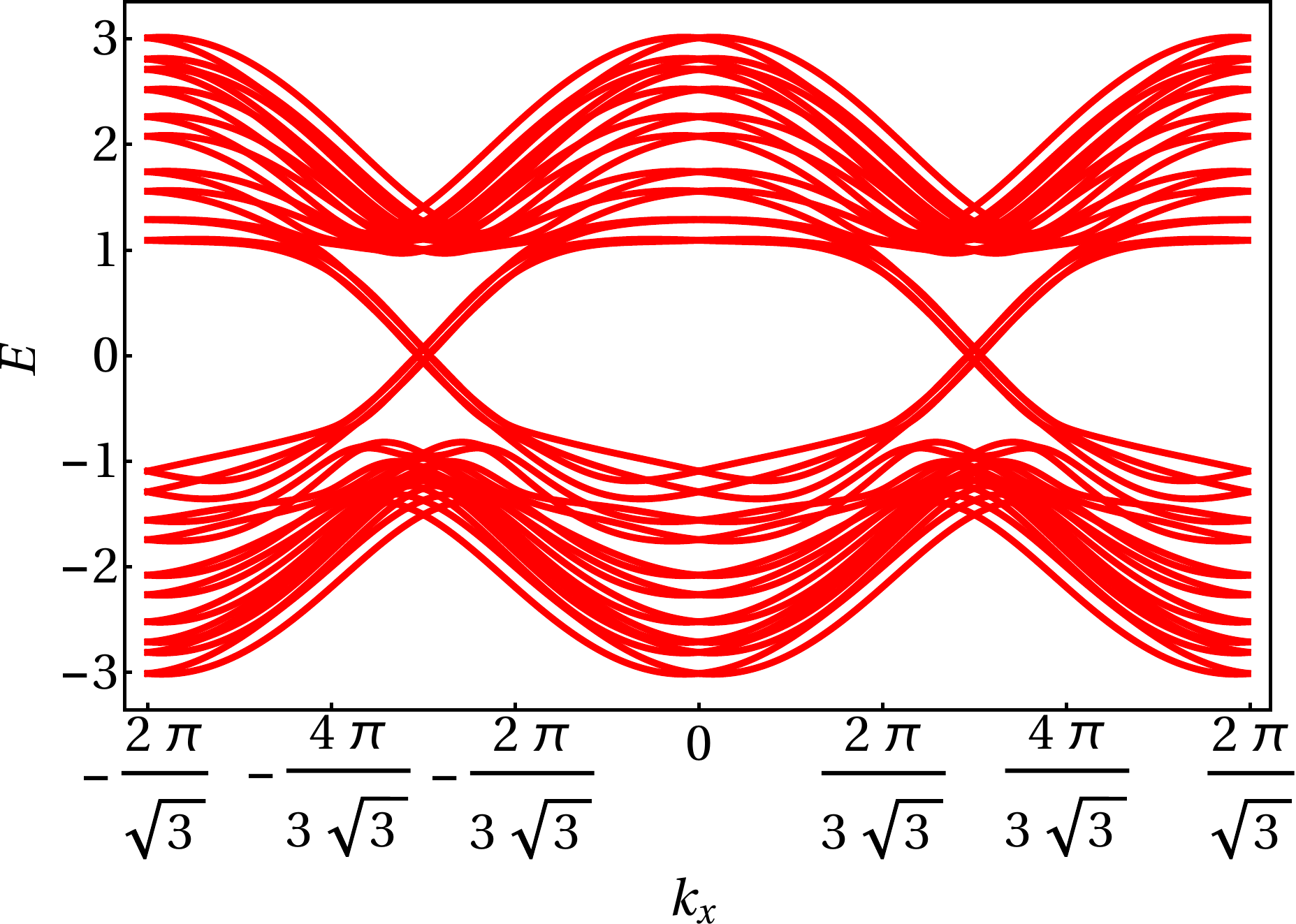}\label{fig:6a}} \hspace*{0.1 cm}
    \subfloat[]{\includegraphics[trim=0 0 0 0,clip,width=0.24\textwidth]{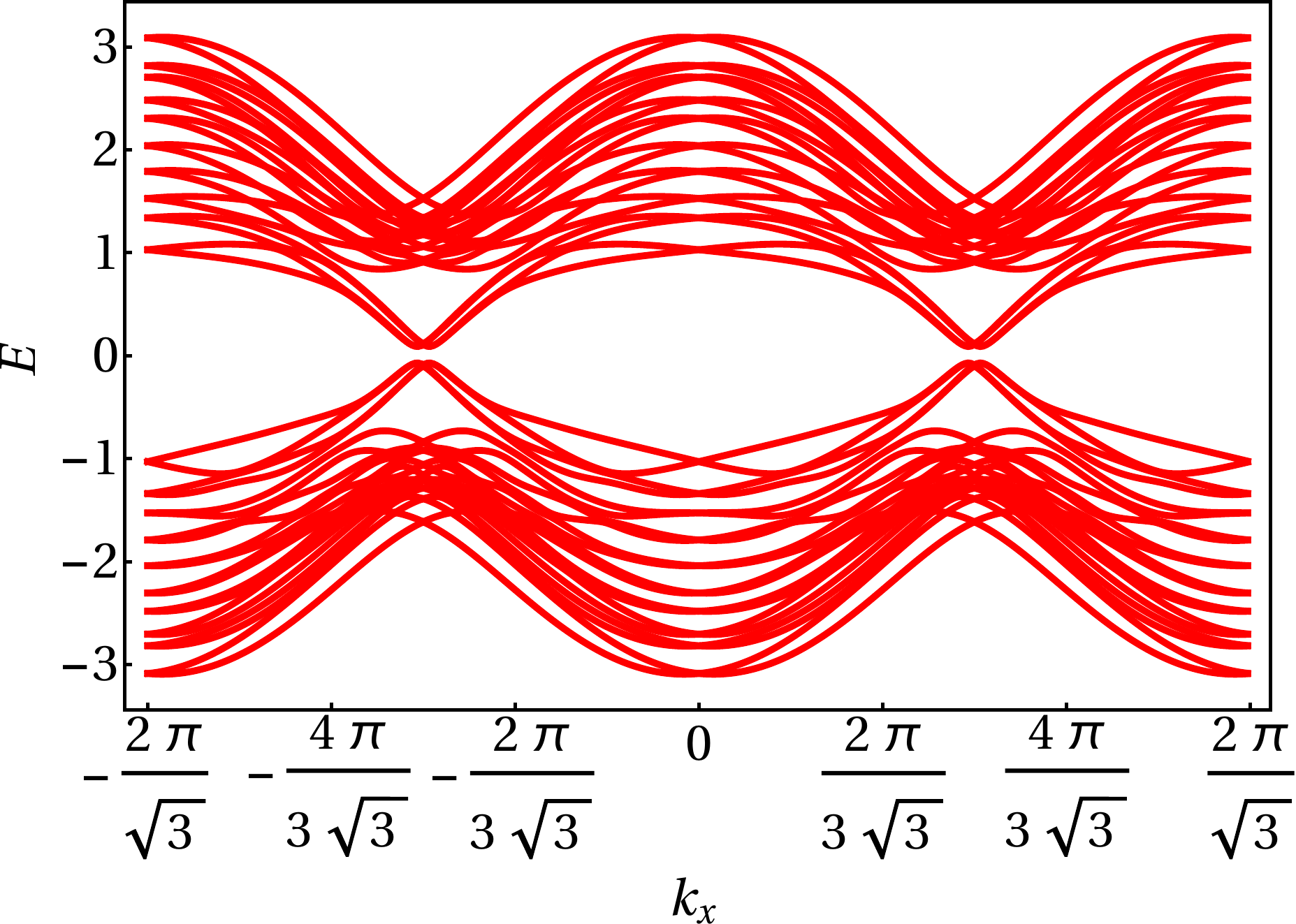}\label{fig:6b}}\\
   \subfloat[]{\includegraphics[trim=0 0 0 0,clip, width=0.24\textwidth]{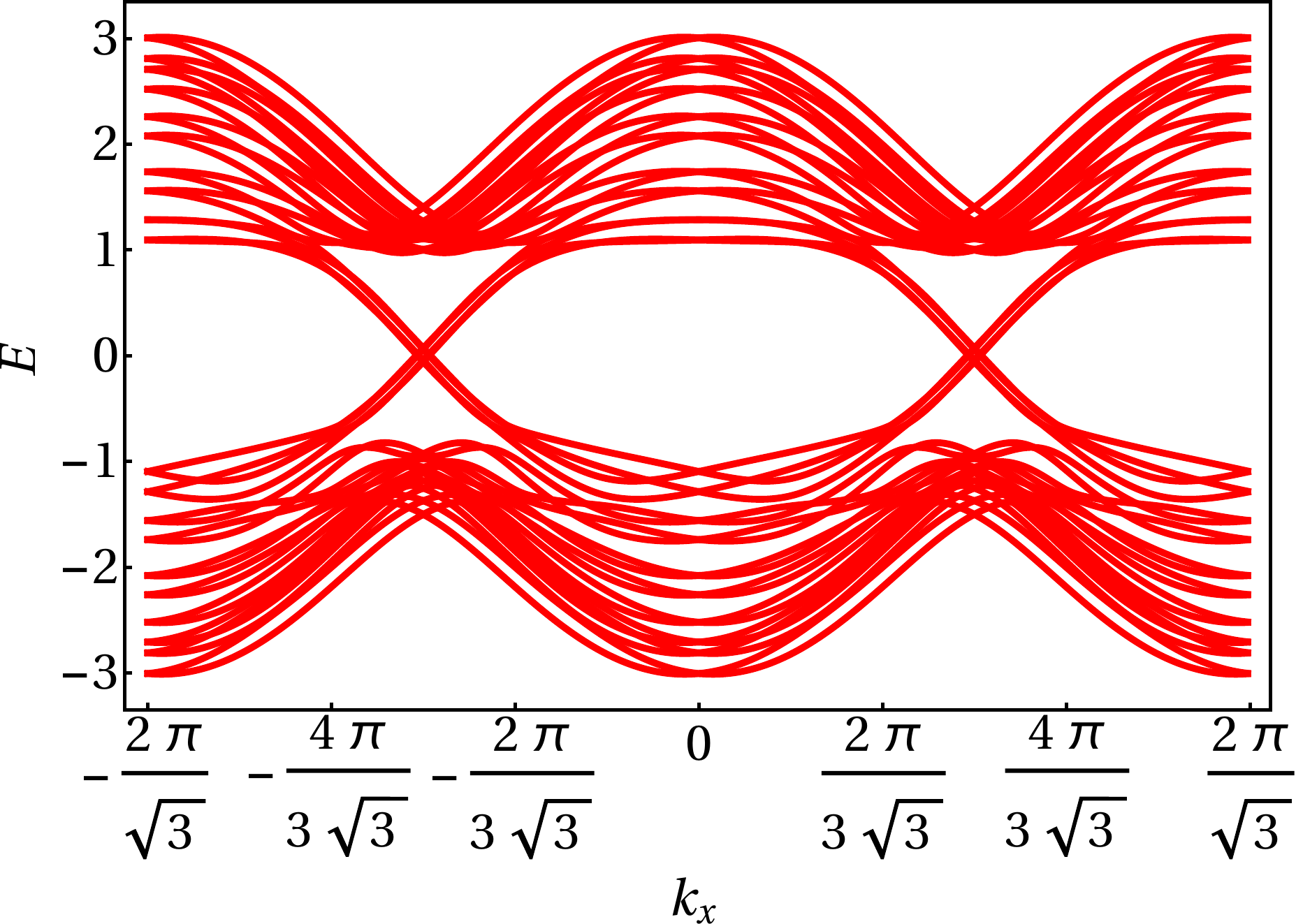}\label{fig:6c}} \hspace*{0.1 cm}
   \subfloat[]{\includegraphics[trim=0 0 0 0,clip,width=0.24\textwidth]{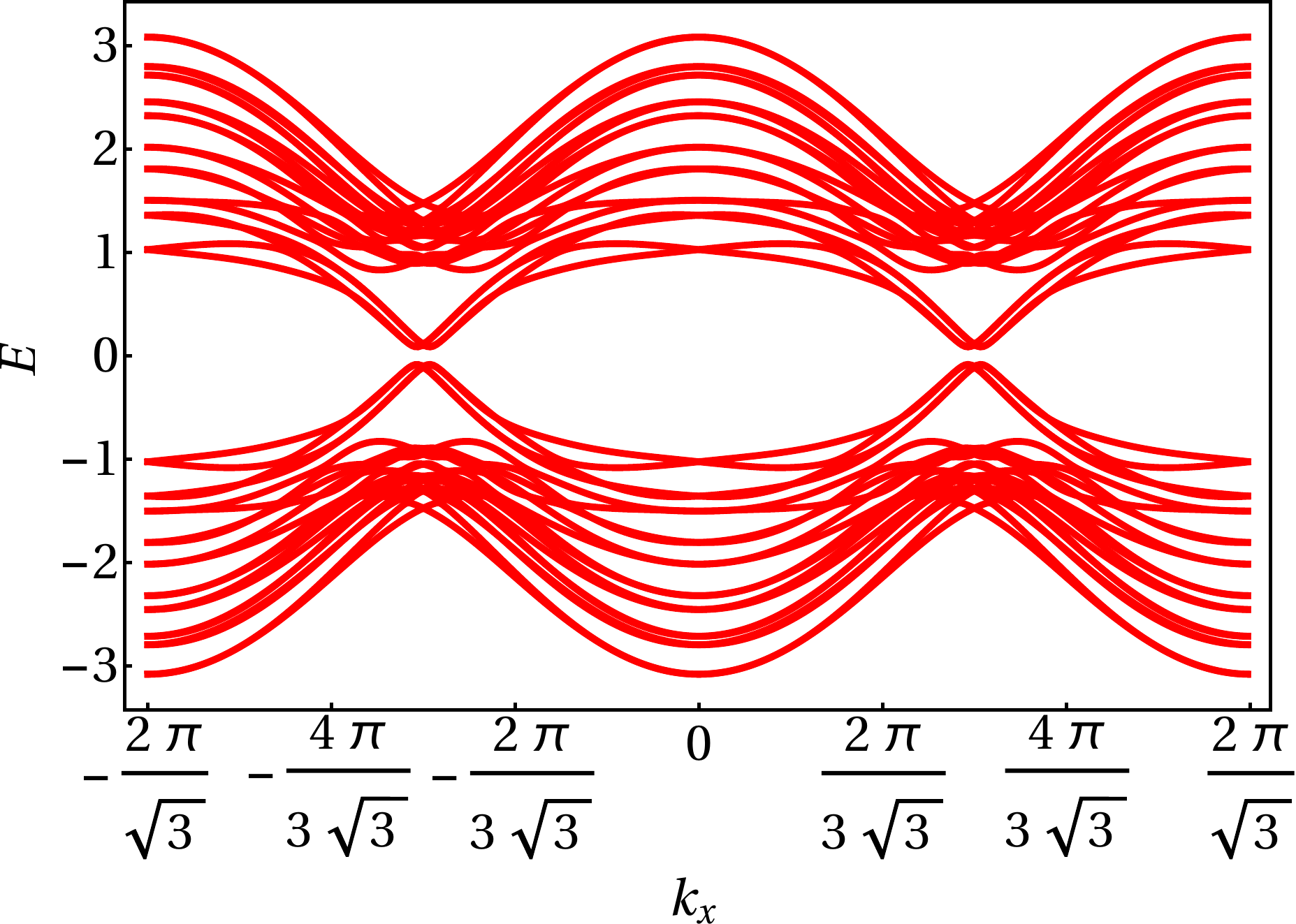}\label{fig:6d}}
 \caption{(color online) The band structure for (a) $\lambda_{R}=0.1$ and $\lambda^{\perp}_{R}=0.05$ (b) $\lambda_{R}=0.1$ and $\lambda^{\perp}_{R}=0.3$ (c) $\lambda_{R}=0.1$ and $\lambda^{\perp}_{R}=0$ (d) $\lambda_{R}=0$ and $\lambda^{\perp}_{R}=0.3$. Here, we set $t_\perp= 0.2$, $t_2= 0.2$ and $N= 5$.}
\label{fig:6}
\end{figure}

\begin{figure}[!ht]
 \centering
   \subfloat[]{\includegraphics[trim=0 0 0 0,clip,width=0.25\textwidth]{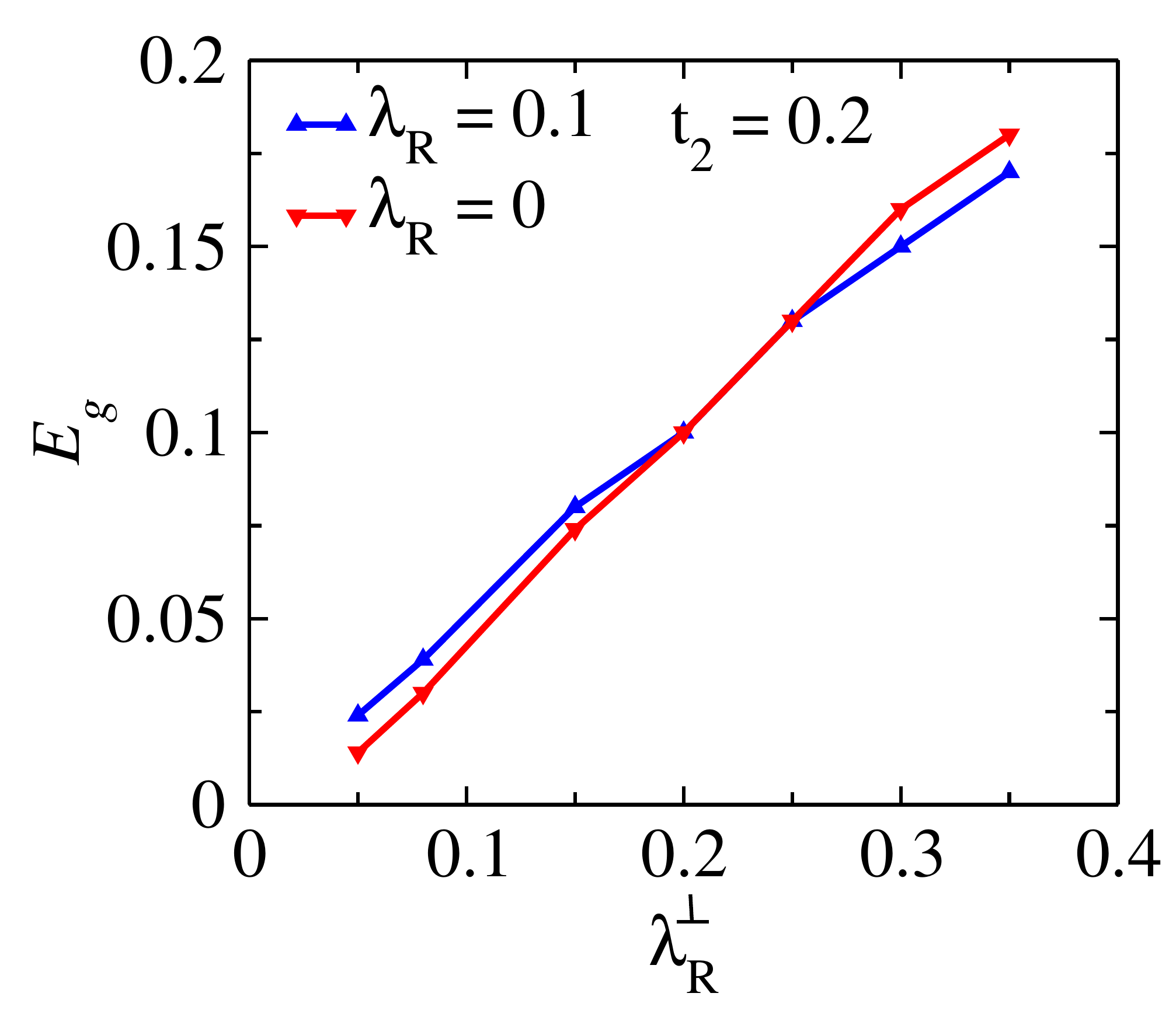}\label{fig:7a}} 
    \subfloat[]{\includegraphics[trim=0 0 0 0,clip,width=0.235\textwidth]{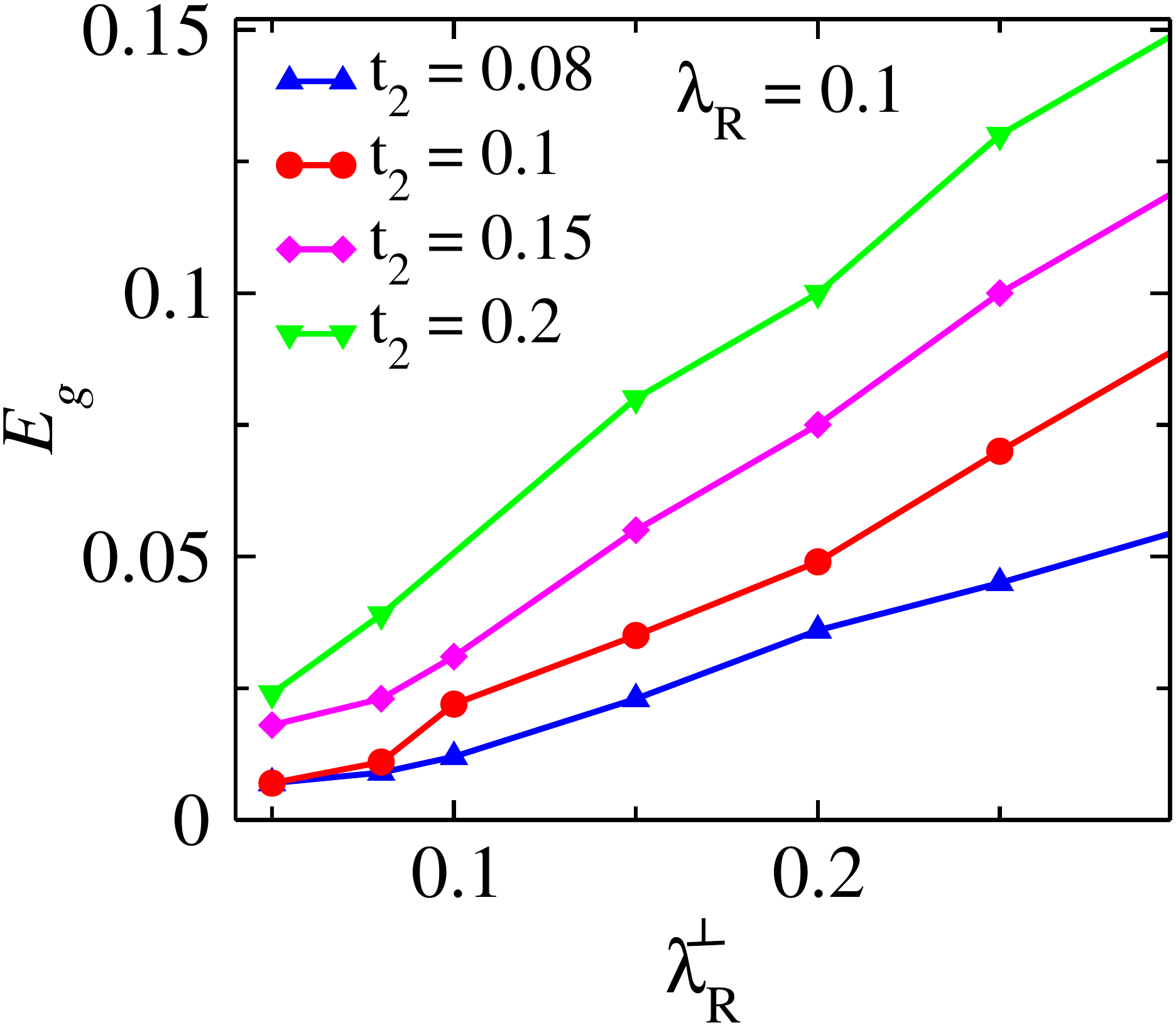}\label{fig:7b}}
  \caption{(color online) Energy gap, $E_g$ as a function of interlayer Rashba coupling parameter, $\lambda^{\perp}_R$ (a) for $\lambda_R=0, 0.1 $ and $t_2= 0.2$ (b) for different values of $t_2$ and $\lambda_R=0.1$. Here, we set $t_\perp=0.2$.}
\label{fig:7}
\end{figure}

\begin{figure*}[!ht]
\centering
   \subfloat[]{\includegraphics[trim=0 0 0 0,clip,width=0.3\textwidth]{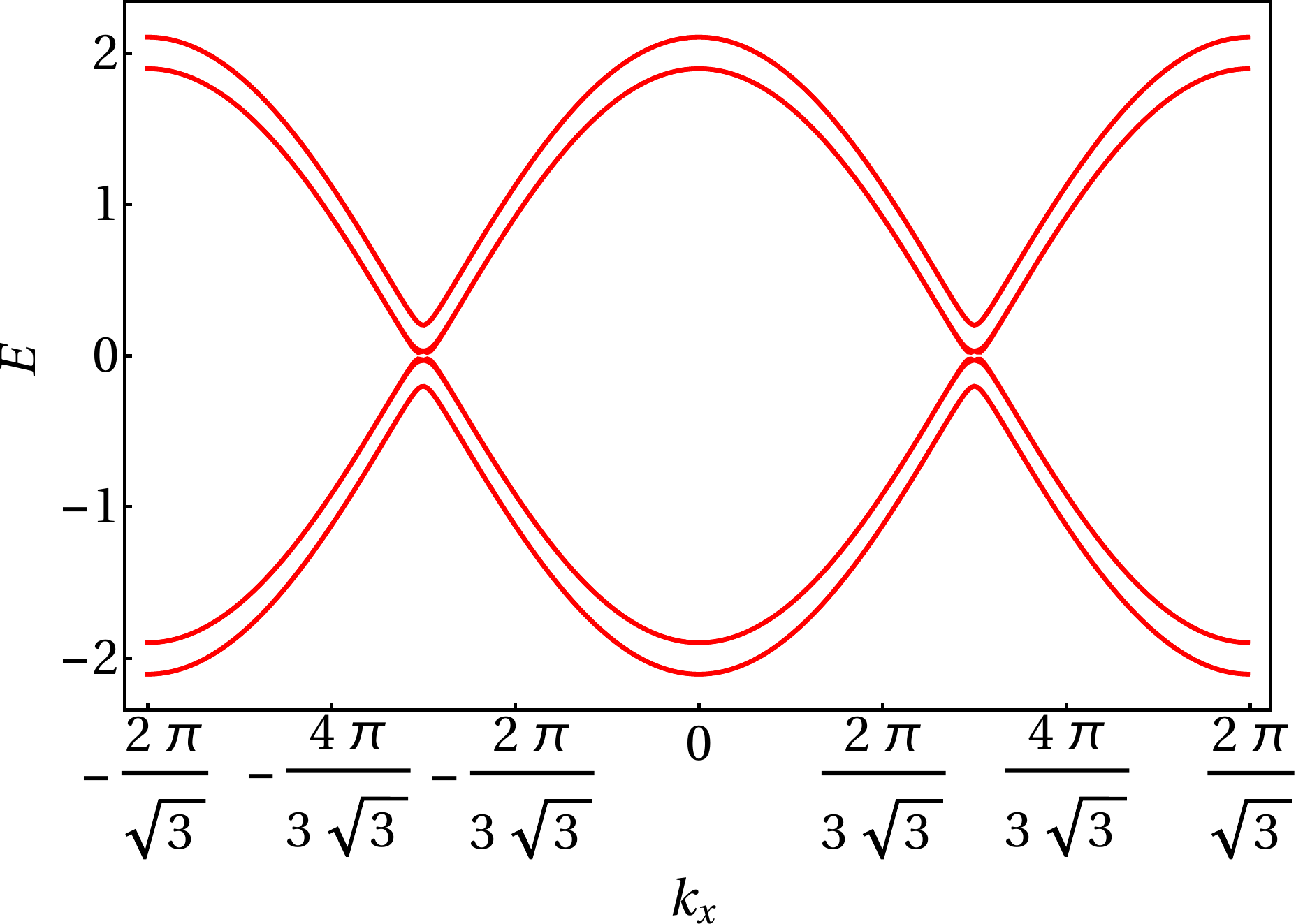}\label{fig:8a}} \hspace*{0.1 cm}
    \subfloat[]{\includegraphics[trim=0 0 0 0,clip,width=0.3\textwidth]{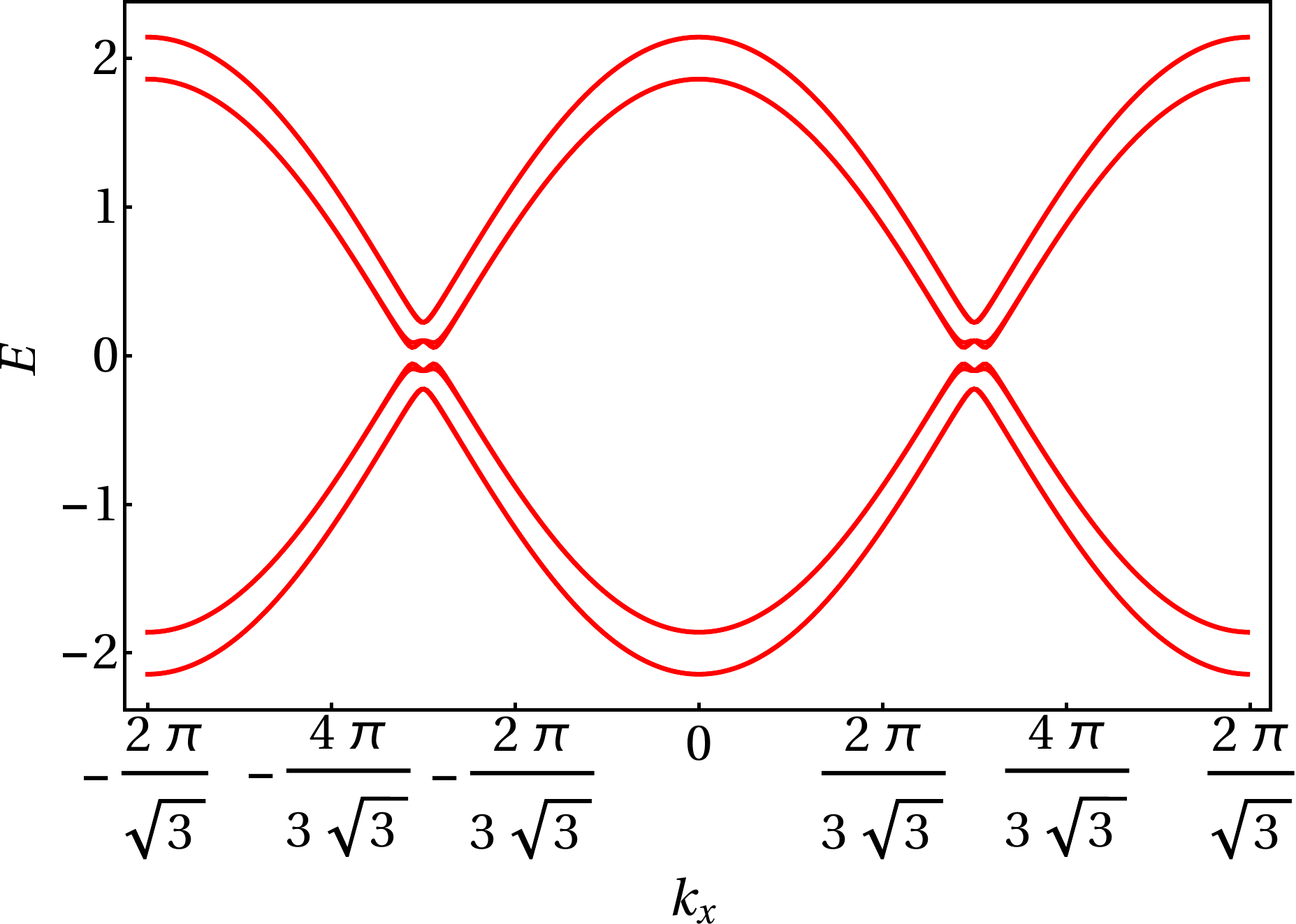}\label{fig:8b}}\hspace*{0.1 cm}
   \subfloat[]{\includegraphics[trim=0 0 0 0,clip,width=0.3\textwidth]{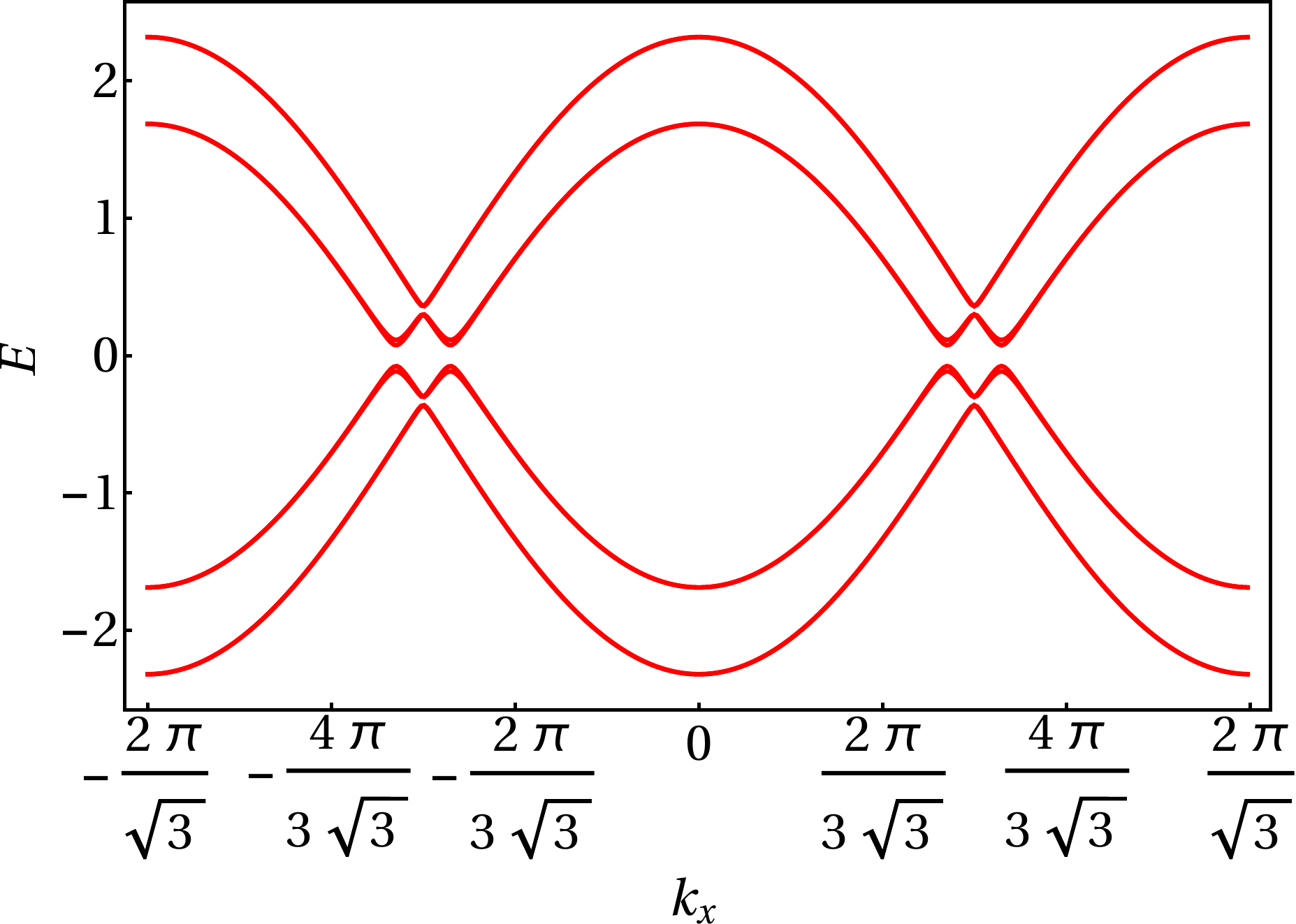}\label{fig:8c}} \\
   \subfloat[]{\includegraphics[trim=0 0 0 0,clip,width=0.3\textwidth]{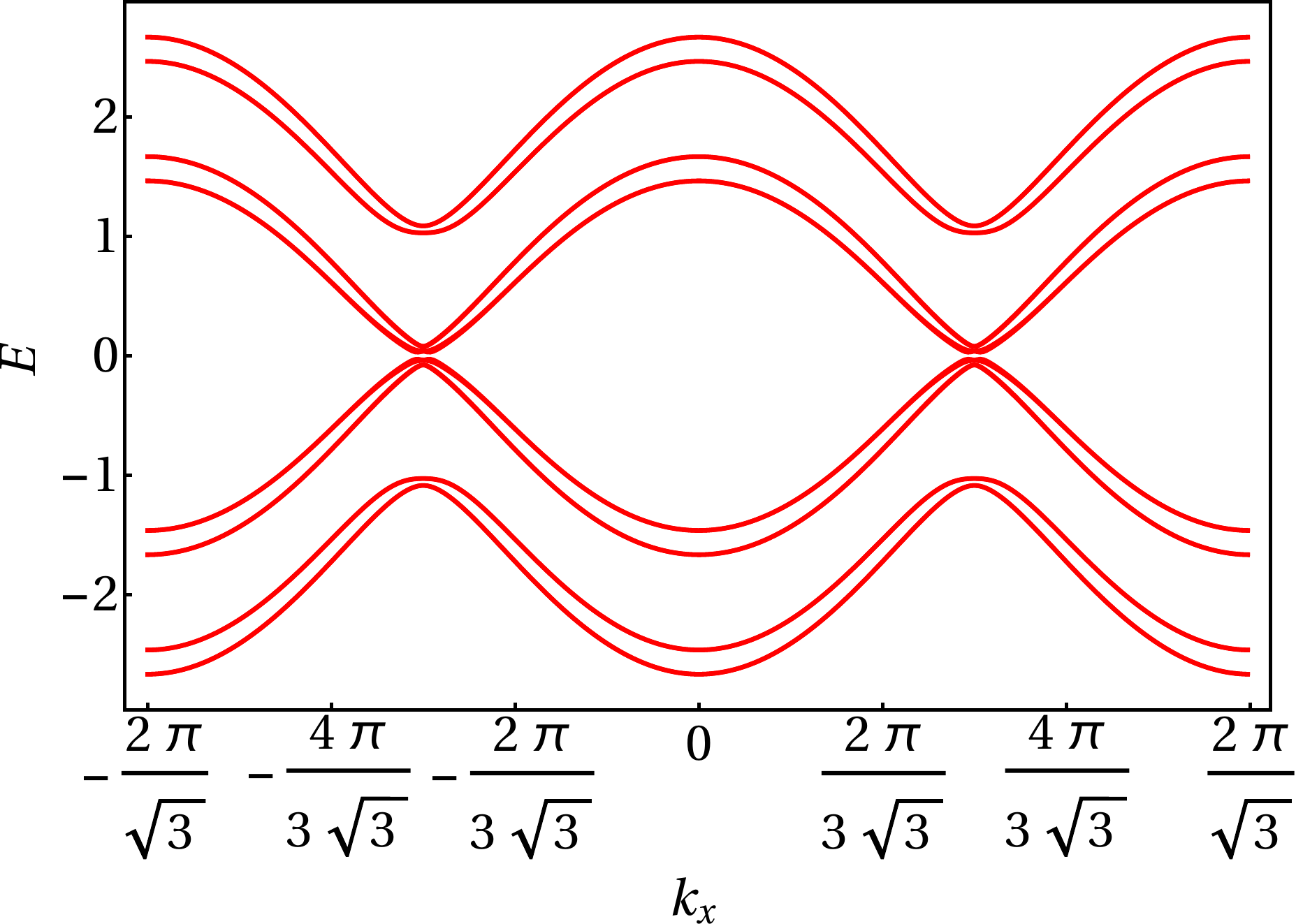}\label{fig:8d}} \hspace*{0.1 cm}
    \subfloat[]{\includegraphics[trim=0 0 0 0,clip,width=0.3\textwidth]{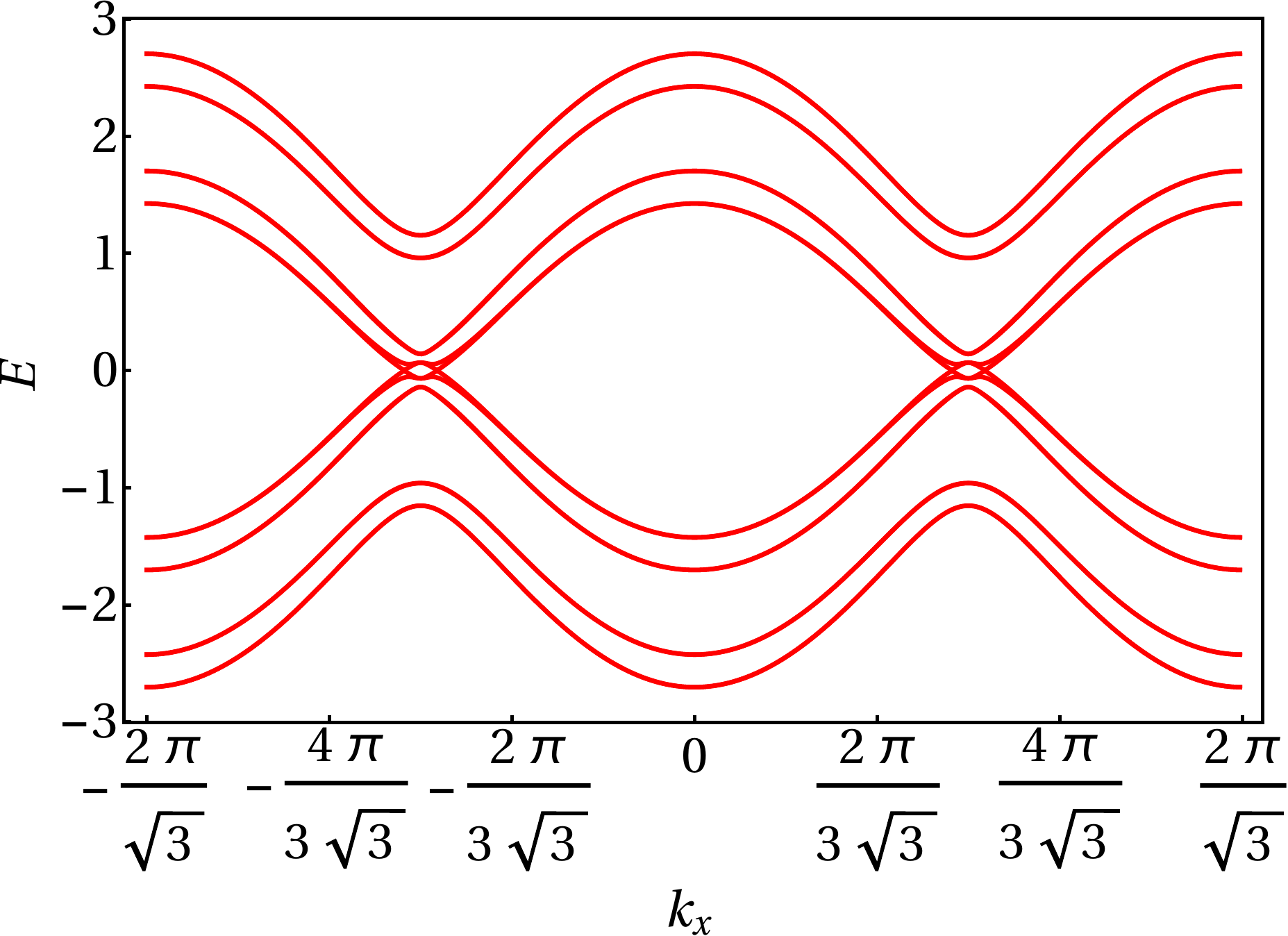}\label{fig:8e}}\hspace*{0.1 cm}
   \subfloat[]{\includegraphics[trim=0 0 0 0,clip,width=0.3\textwidth]{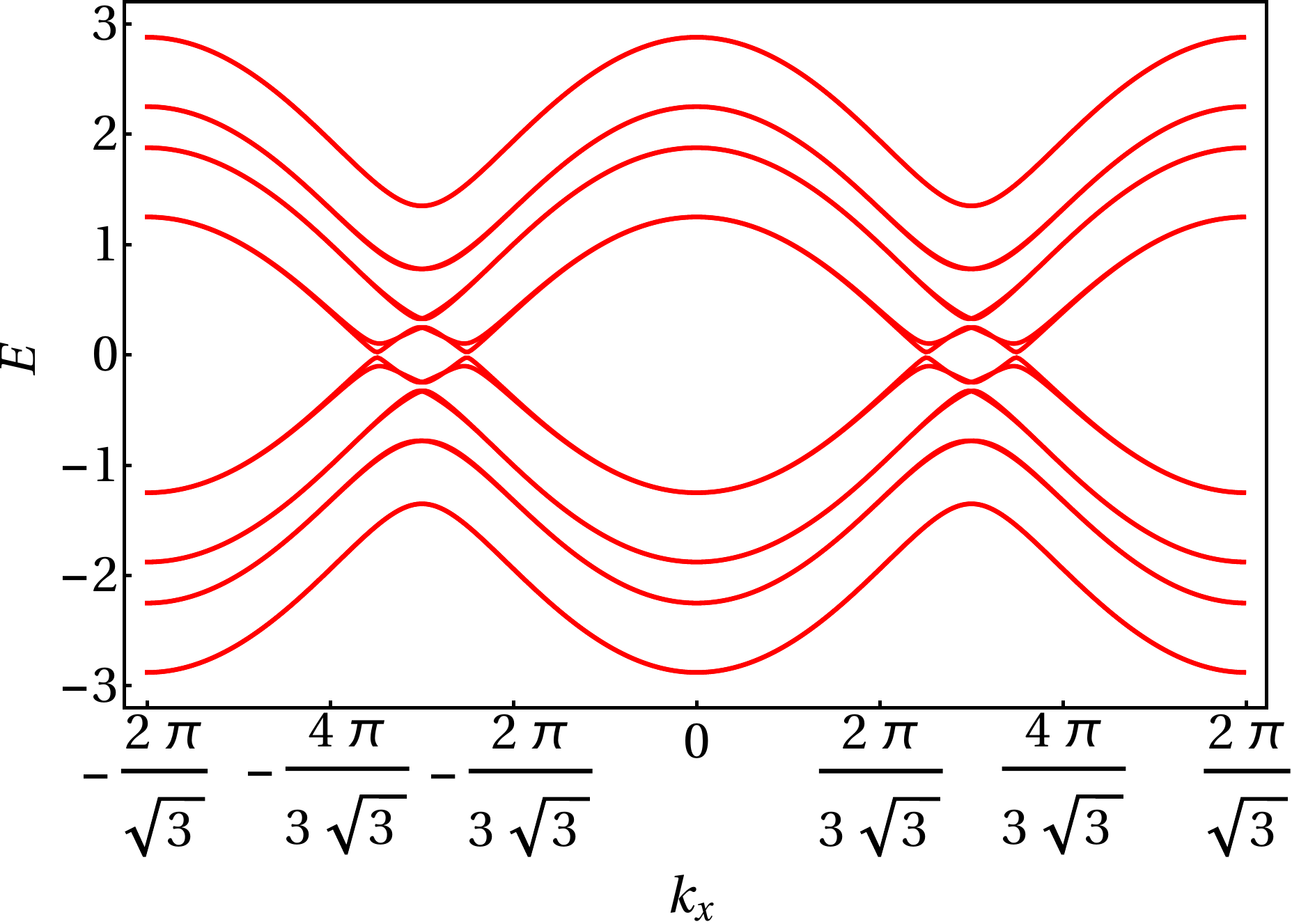}\label{fig:8f}}
   \caption{(color online) The band structure for different values of bias voltage $ V= 0.03, 0.1, 0.3 $ for $N$ = odd (a-c) and $N$ = even (d-f). Here, we set $t_\perp=0.2$ and $t_2= 0.1$. Other parameters, $\lambda_R$ and $\lambda_R^\perp$ both are zero.}
\label{fig:8}
\end{figure*}

Next we have considered intralayer intrinsic SOC and Rashba SOC in and between the layers. Here we have all non-zero values of the coupling parameters, namely,  $t_2\neq 0$, $\lambda_{R}\neq 0$ and $\lambda^{\perp}_{R}\neq 0$. To see the effects of interlayer Rashba coupling, we have fixed the intralayer intrinsic SOC ($t_2=0.2$) and intralayer Rashba SOC ($\lambda_R=0.1$). We have plotted the charge density as shown in Fig.~\ref{fig:5} for two different values of the interlayer Rashba SOC, namely $\lambda^{\perp}_{R}= 0.05$ and $0.3$. It can be seen from Fig.~\ref{fig:5a} that for a small value of $\lambda^{\perp}_{R}$, the amplitudes of the A and B sublattices penetrate somewhat gradually into the bulk for a fixed value of intralayer intrinsic SOC and Rashba SOC, while for large values of $\lambda^{\perp}_{R}$ the penetration depth is enhanced where the fall off of the amplitudes become even more gradual (see Fig.~\ref{fig:5b}). The amplitudes in either of the cases do not exhibit any sharp fall off. \par 
We have also plotted the band structure for the above mentioned values of parameter for $N=5$ (Fig.~\ref{fig:6a} and Fig.~\ref{fig:6b}). It can be seen that for a small value of $ \lambda^{\perp}_{R}$, there exists a band gap which is vanishingly small (shown in Fig.~\ref{fig:6a}) and the spin degeneracy is lifted. However, for large values of $ \lambda^{\perp}_{R}$, the band gap increases and it varies almost linearly with the interlayer Rashba spin orbit coupling. This indicates that the topological properties of QSH phase is destroyed in presence of both intralayer and interlayer Rashba coupling along with intralayer intrinsic SOC. For both the above cases, we have fixed the other parameters $t_\perp= 0.2$, $t_2= 0.2$ and $\lambda_R= 0.1$. It is interesting to note that the edge modes can be seen in presence of intrinsic SOC only, while the inclusion of the interlayer Rashba coupling along with the intralayer Rashba SOC destroy the edge modes.\par
To distinguish the effects of intra and inter layer Rashba SOC on the energy spectrum, we have also plotted  Fig.~\ref{fig:6c} and \ref{fig:6d}. For $\lambda_R= 0.1$ and $\lambda_R^{\perp}= 0$, the crossing of the bands in the $k_x$ range as observed before in presence of only intrinsic SOC remains as it is (as shown in Fig.~\ref{fig:6c}), while for $\lambda_R=0$ and $\lambda_R^{\perp}=0.3$ the band gap opens replacing the scenario of crossing edge modes. For a lucid visualization, we have plotted the band gap, $E_g$ as a function of interlayer Rashba coupling, $\lambda_R^\perp$ for two different cases. Fig.~\ref{fig:7a} shows that although the band gap, $E_g$ varies linearly for the $\lambda_R=0$ and $\lambda_R=0.1$ for a fixed $t_2$, the gap increases more sharply for $\lambda_R=0$ than the other one, while Fig.~\ref{fig:7b} shows the same plot for different values of $t_2$. Thus the energy gap at the Dirac points are affected by the interlayer Rashba SOC in a roughly linear fashion predominantly.  
\subsection{Turning on the bias voltage ($V\neq0)$}
\label{bias}
In this section, we include a biasing term $V$ (such that a constant potential difference $2V$ exists between the layers) and study the effect of this bias voltage in presence of spin-orbit interactions. It is well-known that if we add a bias voltage to bilayer graphene it will open a gap and \textit{mexican-hat} like feature can be observed in the lowest energy band around the Dirac points\cite{castro}. In our previous study, we have seen that there is an odd-even asymmetry in the energy spectrum in presence of intrinsic SOC for very small $N$ with no bias voltage ($V=0$). For $V \neq 0$, the odd-even asymmetry still can be observed in the energy spectrum with a change in the bias voltage. In particular, we have considered three different values of $V$, namely $V=0.03$, $0.1$ and $0.3$. For $N$ = odd, the energy gap in the spectrum increases with increasing bias voltage and the spin degeneracy is lifted only around $k_x= \frac{\pi}{\sqrt3}$ (as shown in Fig.~(\ref{fig:8a}-\ref{fig:8c})). The intrinsic SOC alone cannot result in a spin-lifting spectrum which is observed from Fig.~\ref{fig:4}. The evolution of the \textit{mexican-hat} feature is also observed with increasing bias voltage. The scenario is very different for even values of $N$, for which, when $V$ is small an energy gap is noted with a small magnitude (see Fig.~\ref{fig:8d}). The values of this energy gap is however less than that for $V=0$. With a large value of $V$, namely, $V=0.1$, \textit{mexican-hat} type of feature starts developing with the closing of gap, which eventually becomes more prominent at $V=0.3$ with an energy gap opening at the Dirac points (see Fig.~\ref{fig:8e}). It can be seen that the band transforms from parabolic nature to mexican-hat like nature with the increasing bias voltage.\par
 We also investigate the effect of a non-zero bias voltage of $V$, that is $V\neq0$ with finite $t_2$ and $\lambda_R$. Here we keep $\lambda^{\perp}_R=0$. Without any bias voltage, the intrinsic SOC tends to open a gap in the bulk whereas Rashba SOC tends to close it. If we add a bias voltage the gap increases with the increasing bias voltage for a fixed $t_2$ and $\lambda_R$ (as shown in Fig.~\ref{fig:9}). The particle-hole symmetry no longer exists and the spin degeneracy is lifted except at $k_x=0$ and $\pi/\sqrt3$ due to the Rashba SOC. Intrinsic SOC destroys the mexican-hat like feature and turn it into a parabolic one (not shown in fig), while the \textit{mexican-hat} feature resurfaces with the increasing bias voltage (as shown in Fig.~\ref{fig:9}). We have checked the results with finite $\lambda^{\perp}_{R}$ along with $t_2$ and $\lambda_R$, however no significant change is observed.

\begin{figure}[!ht]
\centering
    \subfloat[]{\includegraphics[trim=0 0 0 0,clip,width=0.24\textwidth]{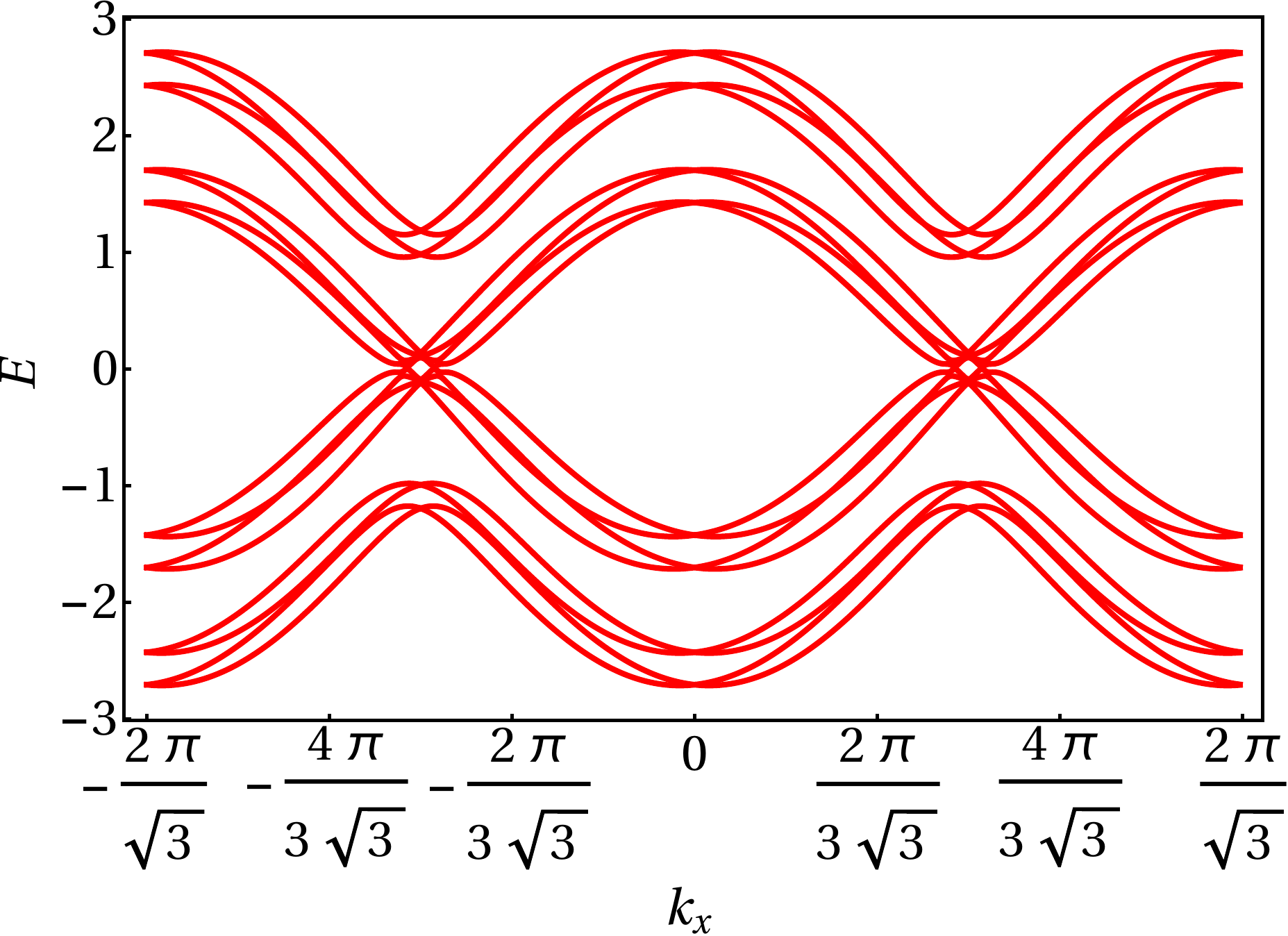}\label{fig:9}}\hspace*{0.1 cm}
   \subfloat[]{\includegraphics[trim=0 0 0 0,clip,width=0.24\textwidth]{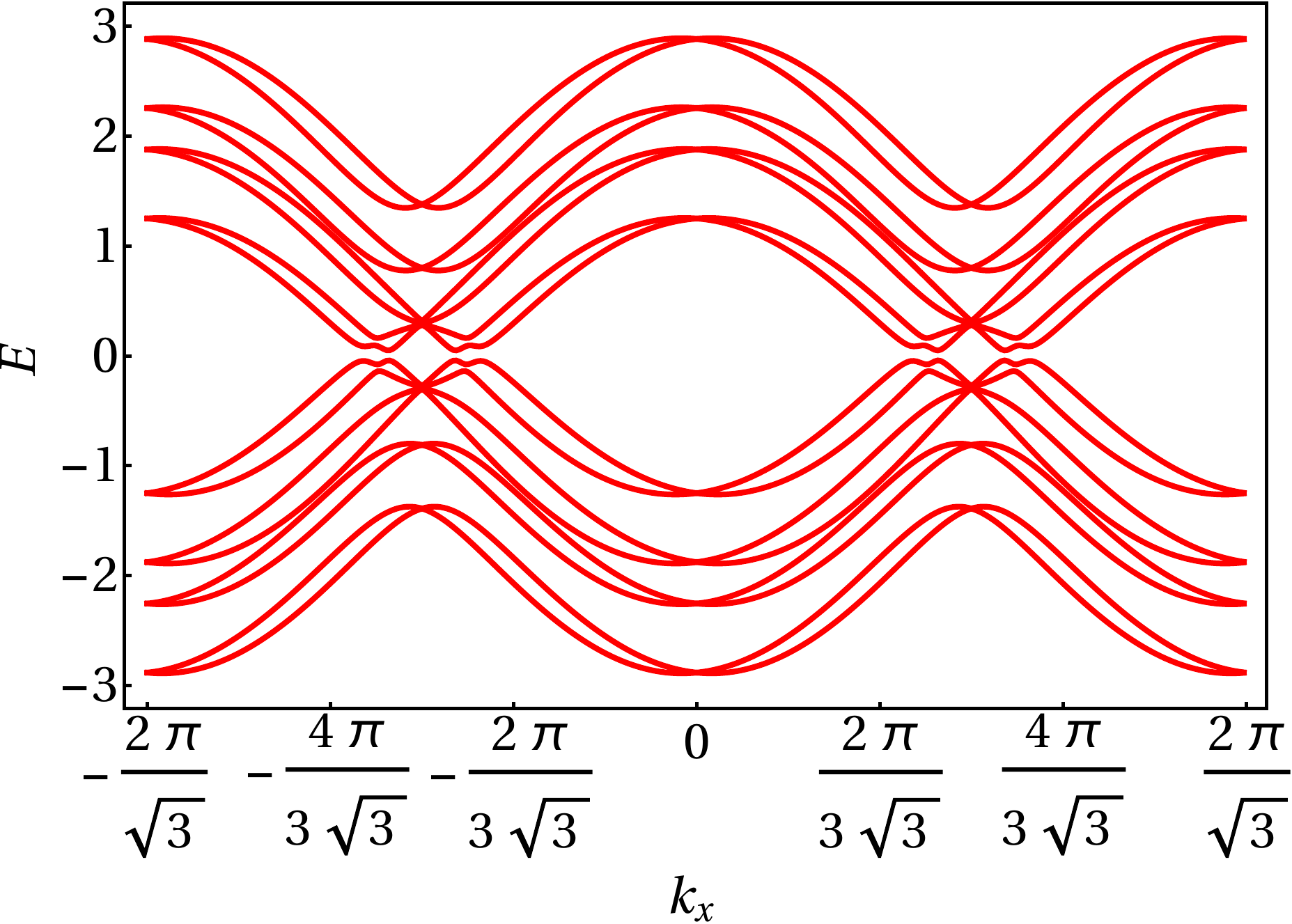}\label{fig:9}}
 \caption{(color online) The band structure for two different values of bias voltage, $V$ (a) $ V= 0.1$ and (b) $V=0.3 $. Here, we set $t_\perp=0.2$, $t_2= 0.1$, $\lambda_R=0.1$ and $\lambda_R^{\perp}=0$.}
\label{fig:9}
\end{figure}

\section{Transport properties}
\label{transport}
The transport properties of a bilayer KM model can be affected by the presence of their edge states. The electron conductance can be calculated using
Landauer-B\"uttiker formula \cite {land_cond,land_cond2} that relates
the zero temperature conductance, $G$ with the transmission coefficient, $T(E)$ as,
\begin{equation}\label{eq_15}
G = \frac{e^2}{h} T(E)
\end{equation}
The transmission
coefficient of a bilayer graphene (as shown in Fig.~\ref{fig:10}) can be calculated via \cite{caroli,Fisher-Lee,dutta},
\begin{equation}\label{eq_16}
T = \text{Tr}\left[\Gamma_R {\cal G}_R
  \Gamma_L {\cal G}_A\right]
\end{equation}

${\cal G}_{R(A)}$ is the retarded (advanced) Green's function corresponding to the
scattering region. The coupling matrices $\Gamma_{L(R)}$ are pertaining to the
imaginary parts representing the coupling between the scattering
region and the left (right) lead. They are defined by \cite{dutta},
\begin{equation}\label{eq_18}
\Gamma_{L(R)} = i\left[\Sigma_{L(R)} -
  (\Sigma_{L(R)})^\dagger\right]
\end{equation}
Here $\Sigma_{L(R)}$ is the retarded self-energy associated with the
left (right) lead. 

Also the spin polarized conductance is defined as\cite{chang},
\begin{equation}\label{eq_20}
G_{\gamma}^{s} =\frac{e^2}{h} \text{Tr}\left[\hat{\sigma_{\gamma}}\Gamma_R {\cal G}_R
  \Gamma_L {\cal G}_A\right]
\end{equation}
 where $\gamma$ = $x, y, z$ and $\sigma$ denote the Pauli matrices.\par
 
 \begin{figure}[!ht]
 \centering
   \subfloat{\includegraphics*[trim=0 0 0 0,clip,width=0.3\textwidth,height=0.2\textwidth]{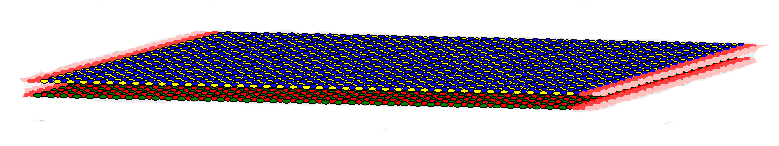}\label{fig:10}}  
  \caption{(color online) Schematic view of zigzag bilayer graphene sheet which consists of a central region, left and right semi-infinite leads (denoted by the red at the both end of the sample). In the central region, blue and yellow circles denote the $A_2$ and $B_2$ sublattices in the upper layer, whereas red and green denotes $A_1$ and $B_1$ sublattices in the lower layer. }
\label{fig:10}
\end{figure}

Next we present our numerical results for the charge and the spin conductances  using Eq.~(\ref{eq_15}-\ref{eq_20}). We have used KWANT\cite{kwant} for our calculations. We have taken $L_x=15$ and $L_y=10$ for zigzag bilayer ribbon to calculate both the charge and spin conductances.

\begin{figure}[!ht]
 \centering \subfloat{\includegraphics[trim=0 0 0
     0,clip,width=0.3\textwidth]{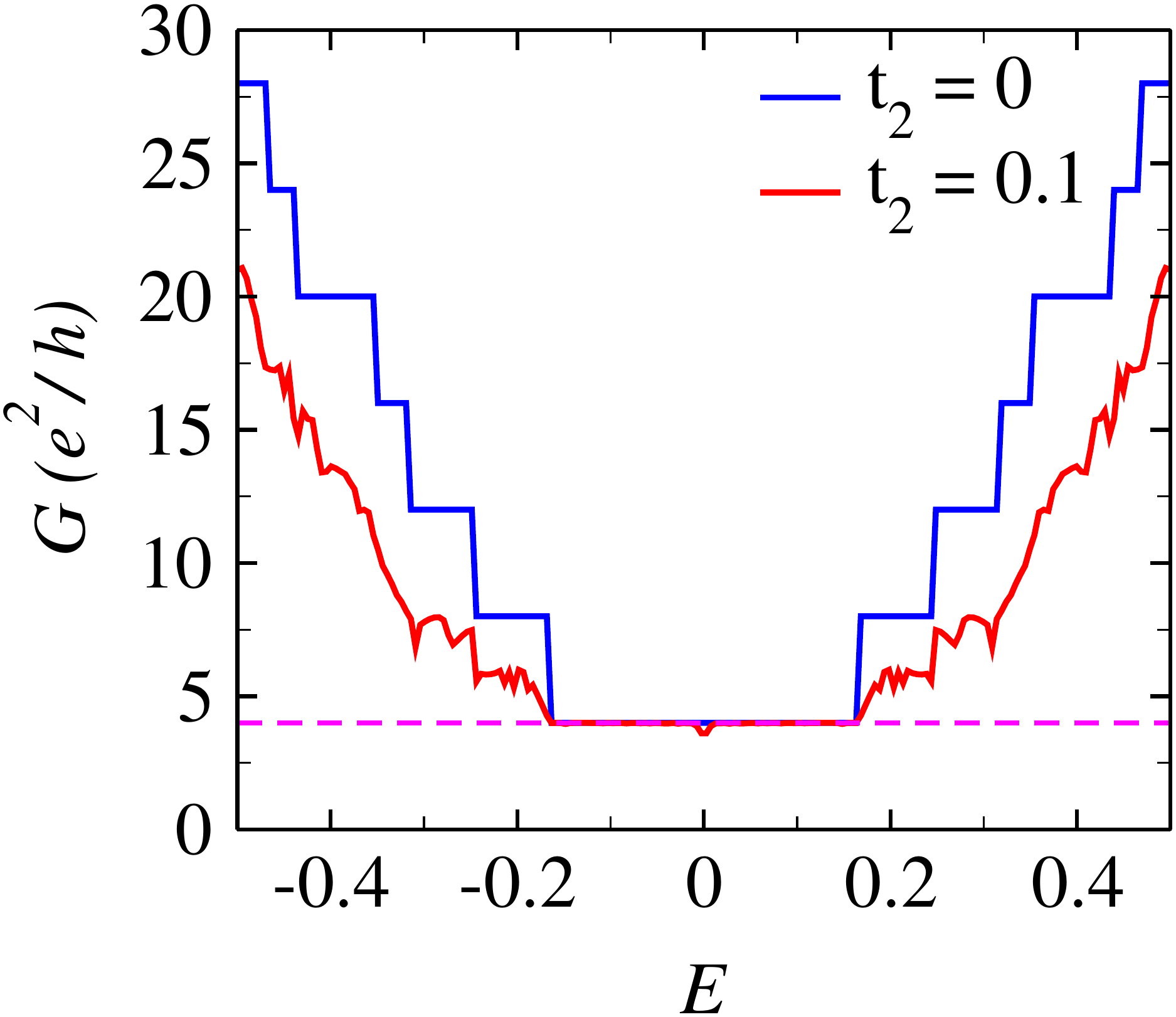}\label{fig:11}}
  \caption{(color online) The charge conductance, $G$ (in units of
    $e^2/h$) as a function of Fermi energy $E$ (in units of $t$) without intralayer intrinsic SOC (the blue curve) and with intralayer intrinsic SOC for $t_2=0.1$ (the red curve). Here, we set $t_\perp=0.2$.}
\label{fig:11}
\end{figure}

We have computed the charge conductance of a bilayer graphene with and without intrinsic SOC as shown in Fig.~\ref{fig:11}. It can be easily observed that for a pristine bilayer though the conductance plot shows step-like behavior emphasizing the basic features of quantum transport phenomena at discrete energy values, the plateau in the vicinity of zero Fermi energy (shown by dotted line in Fig.~\ref{fig:11}) now acquires a value $4e^2/h$ instead of $2e^2/h$ as observed in monolayer graphene. The presence of the edge states is confirmed by the existence of this plateau. The conductance steps are not equidistant along the Fermi energy axis as seen from Fig.~\ref{fig:11} when $t_2 = 0$. The width of these steps depends on the ribbon width and energy intervals\cite{xu}. Further, we show the conductance spectra for $t_2= 0.1$ as shown in Fig.~\ref{fig:11} which records lesser magnitude as compared to a pristine bilayer. Although a plateau is roughly observed at $4e^2/h$, there is an interruption by the presence of the dips around the zero of Fermi energy.

\begin{figure}[!ht]
 \centering
   \subfloat[]{\includegraphics[trim=0 0 0 0,clip, width=0.24\textwidth]{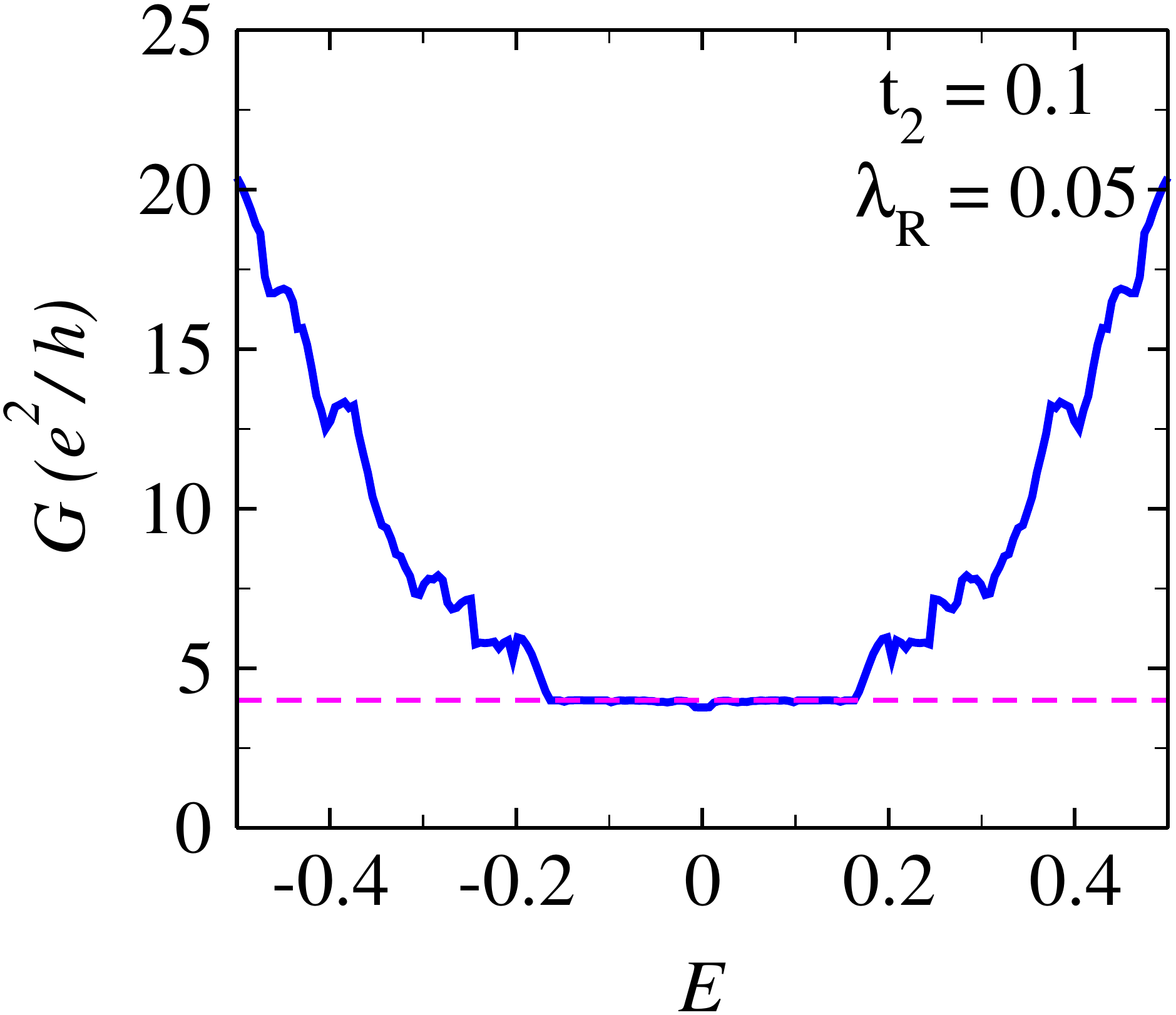}\label{fig:12a}} \hspace*{0.1 cm}
   \subfloat[]{\includegraphics[trim=0 0 0 0,clip,width=0.24\textwidth]{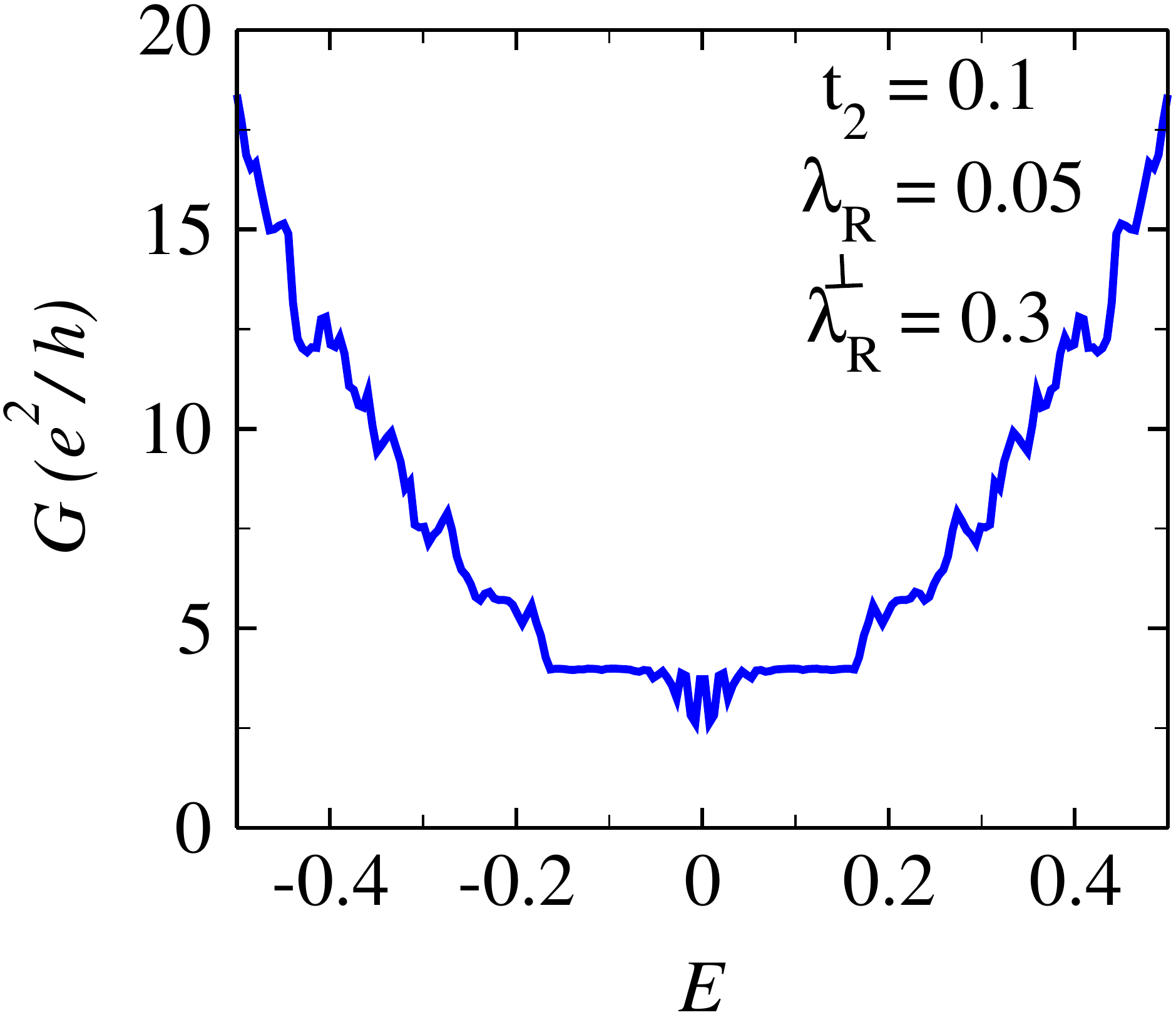}\label{fig:12b}}
   \caption{(color online) The charge conductance, $G$ (in units of
    $e^2/h$) as a function of Fermi energy $E$ (in units of $t$)
    for (a) $\lambda^{\perp}_{R}=0$ (b) $\lambda^{\perp}_{R}=0.3$. Here, we set $t_\perp= 0.2$, $t_2= 0.1$ and $\lambda_R=0.05$.}
\label{fig:12}
\end{figure} 

\begin{figure}[!ht]
 \centering
   \subfloat{\includegraphics[trim=0 0 0 0,clip, width=0.3\textwidth]{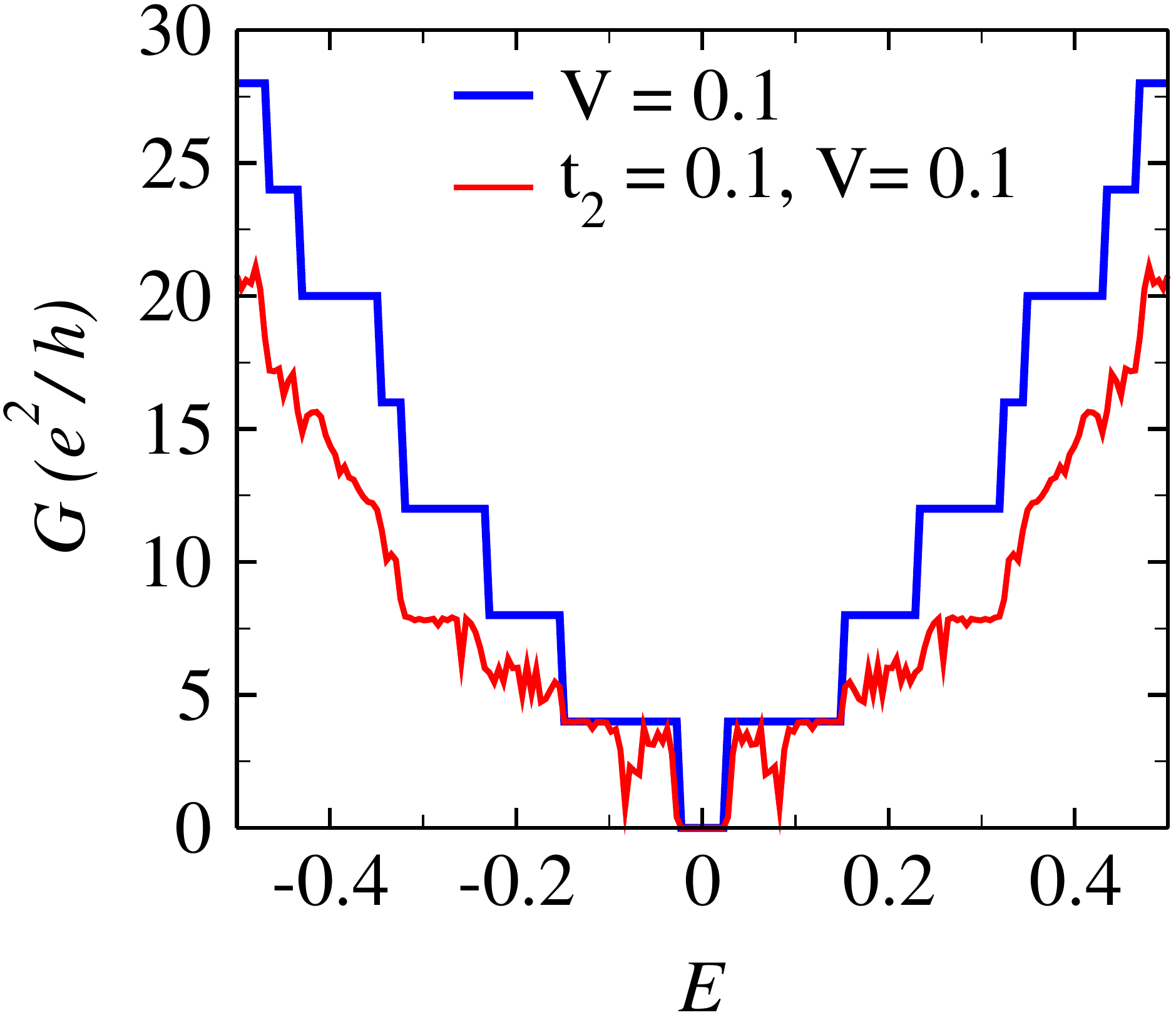}\label{fig:13}}\hspace*{0.1 cm}
   
 \caption{(color online) The charge conductance, $G$ (in units of $e^2/h$) as a function of Fermi energy $E$ (in units of $t$)
    for (a) $V$ = 0.1 (blue curve) and $t_2$ = 0.1, $V$ = 0.1 (red curve) (b) $t_2$ = 0.1, $\lambda_R$ = 0.1, $V$ = 0.1. Here we set  $t_\perp= 0.2$.}
\label{fig:13}
\end{figure}

We have also computed the charge conductance in presence of intralayer Rashba coupling and both intra and interlayer Rashba coupling as shown in Fig.~\ref{fig:12a} and Fig.~\ref{fig:12b} respectively. Fig.~\ref{fig:12a} confirms the existence of edge modes (as shown by the dotted line), whereas Fig.~\ref{fig:12b} shows that the dip is quite sharp around zero Fermi energy and confirms the absence of edge modes as seen from Fig.~\ref{fig:5}.\par

To see the effect of a bias voltage on the conductance properties we have included a potential +$V$ on the upper layer and -$V$ on the lower layer as given in Eq. \ref{eq_1}. Fig.~\ref{fig:13} shows the charge conductance as a function of Fermi energy for a biased bilayer where $V = 0.1$. The blue and red curves correspond to biased bilayer graphene in absence and presence of the intrinsic SOC respectively. As compared to Fig.~\ref{fig:11}, the $4e^{2}/h$ plateau near the zero Fermi energy turns out to zero for both the cases. The conductance steps still remains unaltered but the width of the plateau becomes smaller (as shown in Fig.~\ref{fig:13} by the blue and red curve). The value of the conductance becomes zero due to the opening of the gap between the valence band and conduction band.

\begin{figure*}[!ht]
 \centering
   \subfloat[]{\includegraphics[trim=0 0 0 0,clip, width=0.26\textwidth]{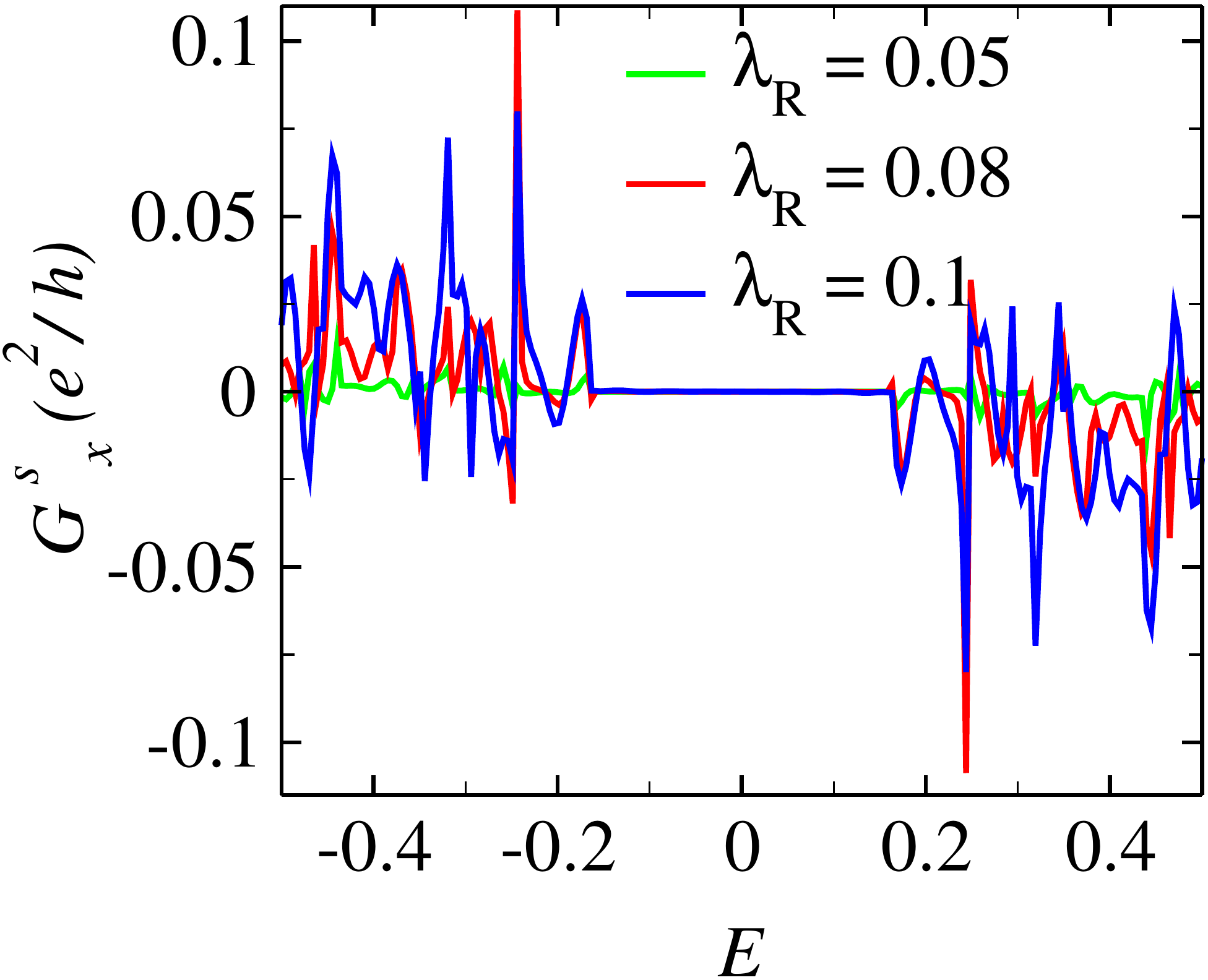}\label{fig:14a}}\hspace*{0.1 cm}
   \subfloat[]{\includegraphics[trim=0 0 0 0,clip,width=0.25\textwidth]{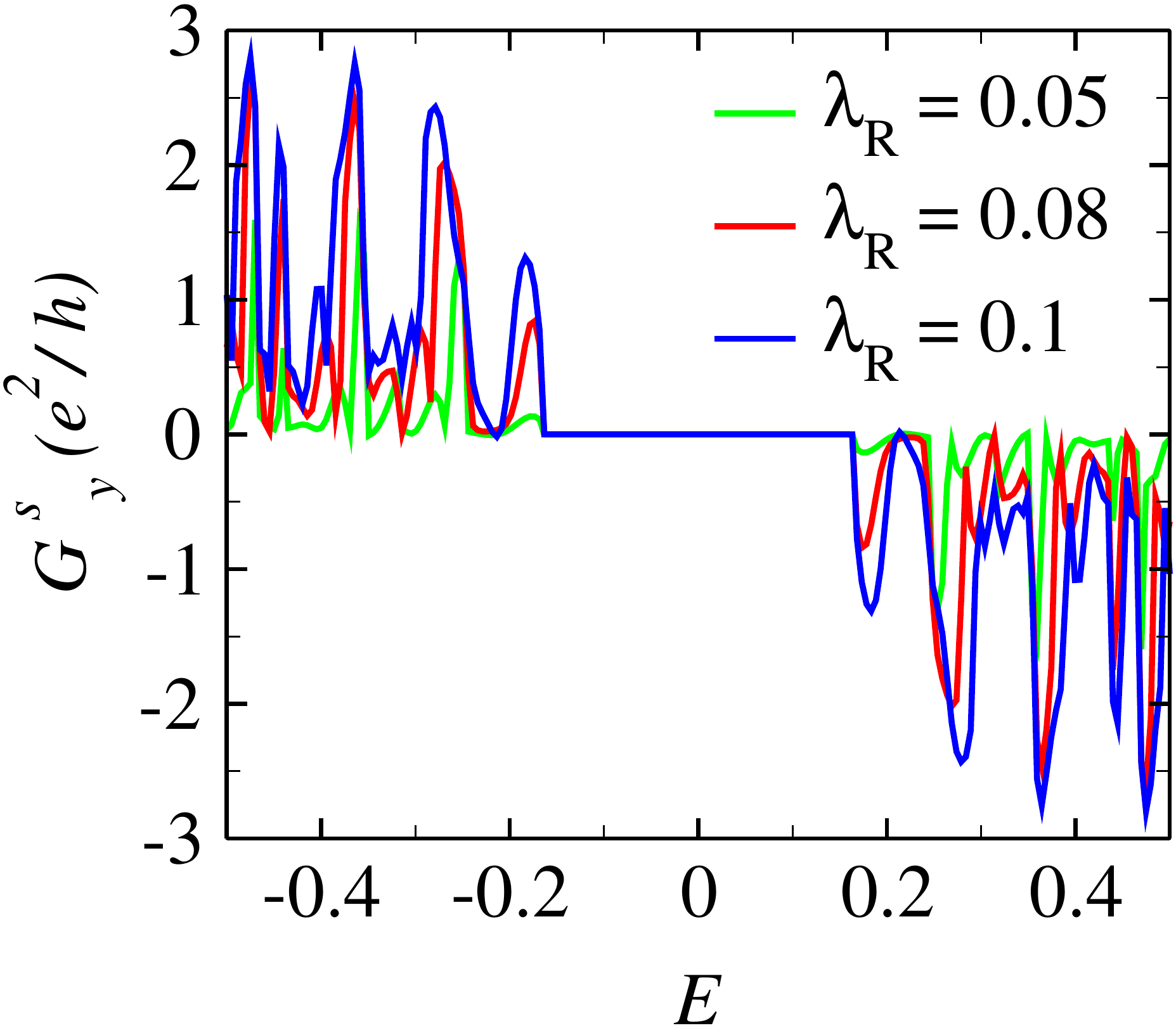}\label{fig:14b}}\hspace*{0.1 cm}
 \subfloat[]{\includegraphics[trim=0 0 0 0,clip,width=0.26\textwidth]{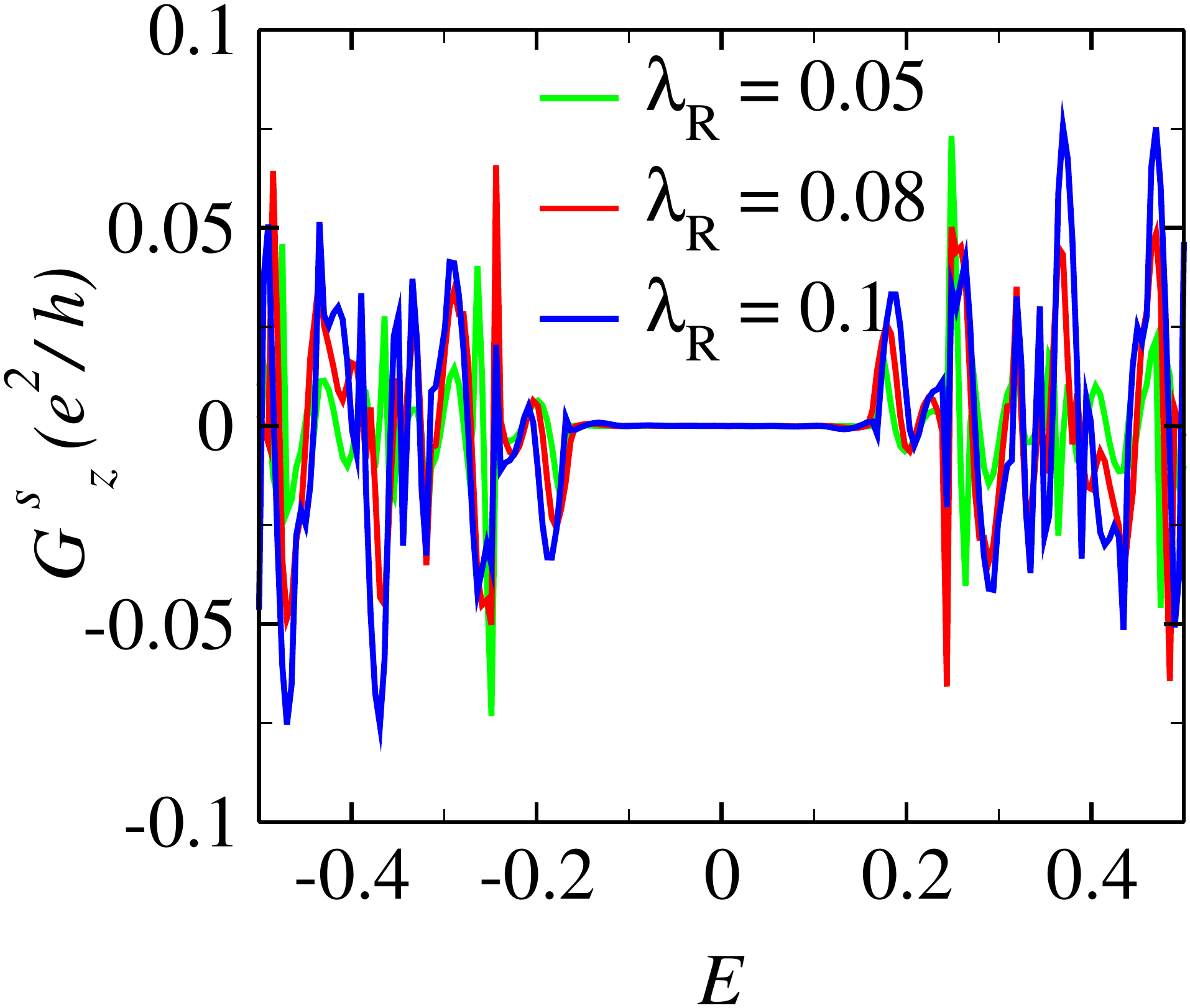}\label{fig:14c}} \\
 \subfloat[]{\includegraphics[trim=0 0 0 0,clip, width=0.26\textwidth]{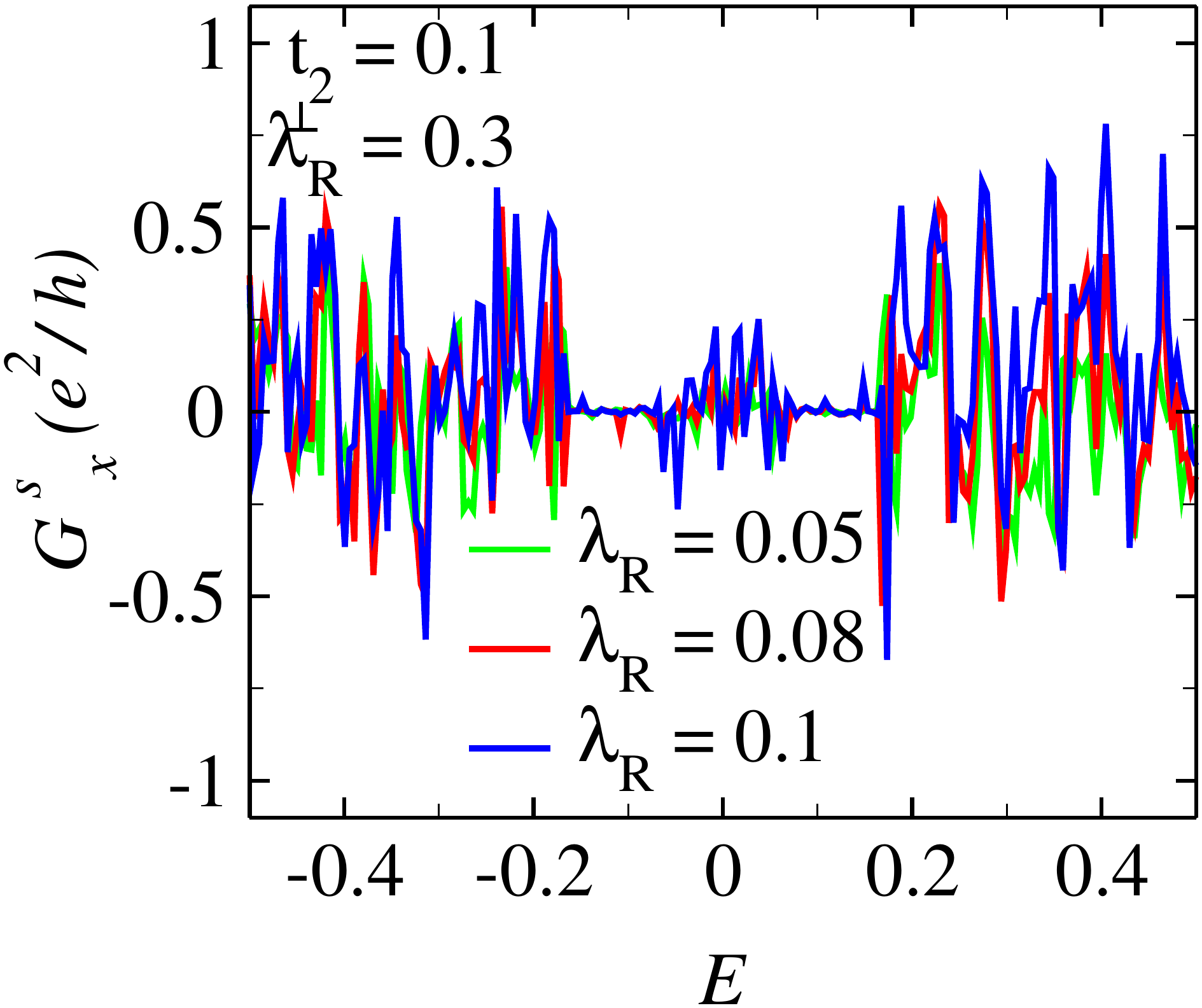}\label{fig:14d}} \hspace*{0.1 cm}
  \subfloat[]{\includegraphics[trim=0 0 0 0,clip, width=0.26\textwidth]{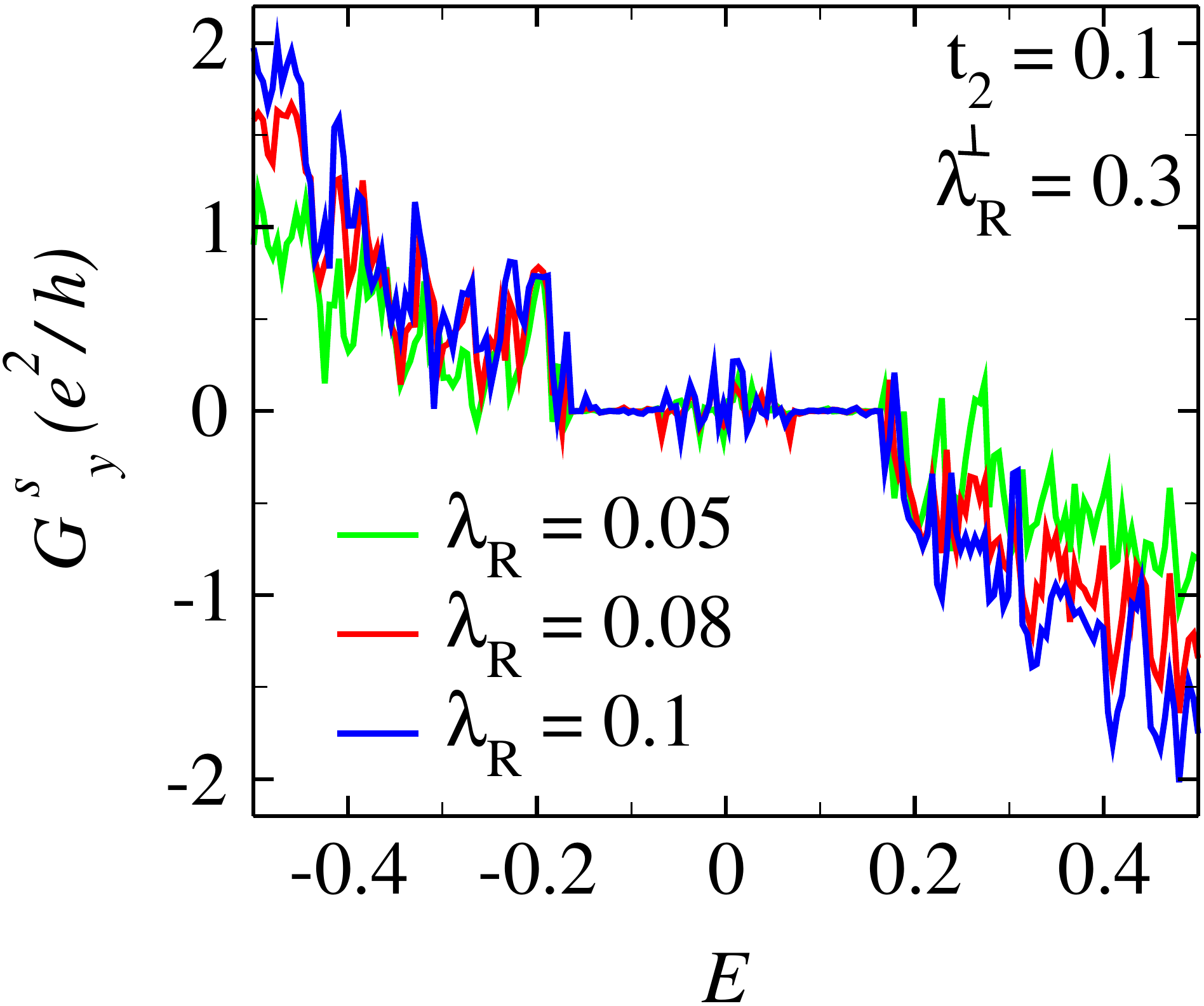}\label{fig:14e}} \hspace*{0.1 cm}
     \subfloat[]{\includegraphics[trim=0 0 0 0,clip, width=0.26\textwidth]{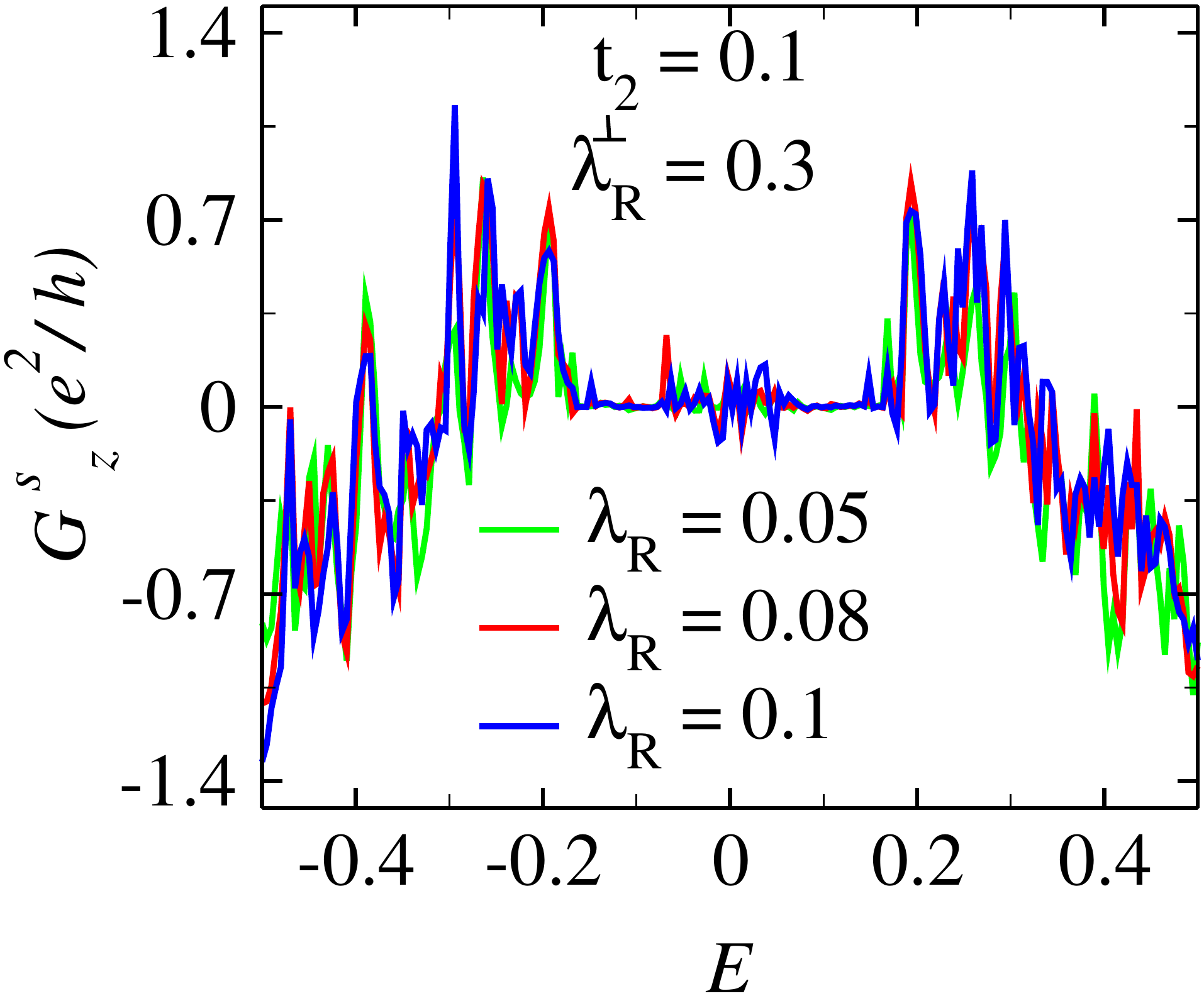}\label{fig:14f}} 
  \caption{(color online) The spin polarized conductance, $G_{\gamma}^{s}$ ($\gamma= x, y, z$) (in units of
    $e^2/h$) as a function of Fermi energy $E$ (in units of $t$)
    for (a-c) $\lambda_R=0.05$, $0.08$ and $0.1$ (Other parameter $t_\perp= 0.2$, $t_2= 0$ and $\lambda_R^{\perp}= 0$) and (d-f) $\lambda_R=0.05$, $0.08$ and $0.1 $ (Other parameter $t_\perp= 0.2$, $t_2= 0.1$ and $\lambda_R^{\perp}= 0.3$).}
\label{fig:14}
\end{figure*}

We further investigated the spin-polarized transport in a zigzag bilayer graphene. Spin polarized conductance results due to the presence of Rashba SOC. It was also shown by Zhang et al\cite{chan, Lin} that the spin polarization components for $x$ and $z$ are zero for an ideal graphene nanoribbon because of the longitudinal mirror symmetry of an infinite system. The spin polarization component corresponding to the $y$-direction is finite and is found to be around $40 \%$ polarized for both the armchair and the zigzag nanoribbon \cite{chan}. However, for a bilayer graphene, we get finite values for all the components of spin polarization. This ensures larger net spin polarized conductance than that of a single layer and hence a bilayer graphene should be a more efficient candidate for spintronic applications. All the three components of the spin polarization, namely, $G_{x}^{s}$, $G_{y}^{s}$ and $G_{z}^{s}$ are plotted as a function of Fermi energy in Fig.~\ref{fig:14a}, \ref{fig:14b} and \ref{fig:14c} for three different values of $\lambda_R$, namely, $0.05$, $0.08$ and $0.1$ with $t_2=\lambda^{\perp}_R=0$. The magnitudes of the $x$ and $z$ component are one order smaller than that of the $y$ component. It can be seen that the spin polarization vanishes in the low energy range and is anti-symmetric in nature around the Fermi energy for all the three components due to electron-hole symmetry present in the Hamiltonian. The nature of the spin polarization for different values of $\lambda_R$ qualitatively remains same but the magnitude gets larger as we increase the strength of the intralayer Rashba parameter, $\lambda_R$.  
In addition, we have included both intralayer intrinsic SOC and interlayer Rashba SOC in our model and have plotted the three components of the spin polarization for three different values of $\lambda_R$ as a function of energy as shown in Fig.~\ref{fig:14d}, \ref{fig:14e} and \ref{fig:14f} corresponding to fixed values of the other spin-orbit coupling parameters, namely, $t_2= 0.1$, $\lambda^{\perp}_R= 0.3$. Though the magnitude of $y$ component of the spin polarization gets smaller (as shown in Fig.~\ref{fig:14e}),  the magnitudes of $x$ and $z$ components grow larger (as shown in Fig.~\ref{fig:14d} and Fig.~\ref{fig:14f}) compared to a scenario where only intralayer Rashba SOC ($\lambda_R$) exists (Fig.~\ref{fig:14a}-\ref{fig:14c}). Importantly, the spin polarization is finite for $E\simeq 0$. The $y$ component is anti-symmetric as a function of Fermi energy $E$, while the same does not hold for the $x$ and $z$ components.

\section{Connection with experiments}
\label{effective_mass}
In order to connect with the experimental data, we have computed the effective mass, $m^{*}$ and investigated its variation with the intralayer Rashba spin-orbit coupling parameter, $\lambda_R$ for a bilayer graphene. The definition of the effective mass and the group velocity can be given as, $m^{*}= \hbar^2k/[dE(k)/dk]$ and $v_g=\frac{1}{\hbar}\frac{dE}{dk}$ (where $k$ is the crystal momentum). Thus $m^{*}=\hbar k/v_{g}$ and hence depends upon the slope of the dispersion near the Dirac points. The measurement of effective mass $m^*$ for a large range of carrier densities in a bilayer graphene was performed using Shubnikov-de Haas (SdH) oscillations \cite{zou}. From the first principles study, it was reported that the effective mass in bilayer graphene is approaximately $0.022 m_e$ ($m_e$ being the bare mass) and also the value increases with the increase in number of layers \cite{srivastava}. Alternatively, the carrier transport properties in a bilayer graphene can also be tuned by the presence of Rashba SOC which could be enhanced by metal-atom adsorption or using an external gate voltage.\par

\begin{figure}[!ht]
 \centering 
  \subfloat{\includegraphics[trim=0 0 0 0,clip, width=0.3\textwidth]{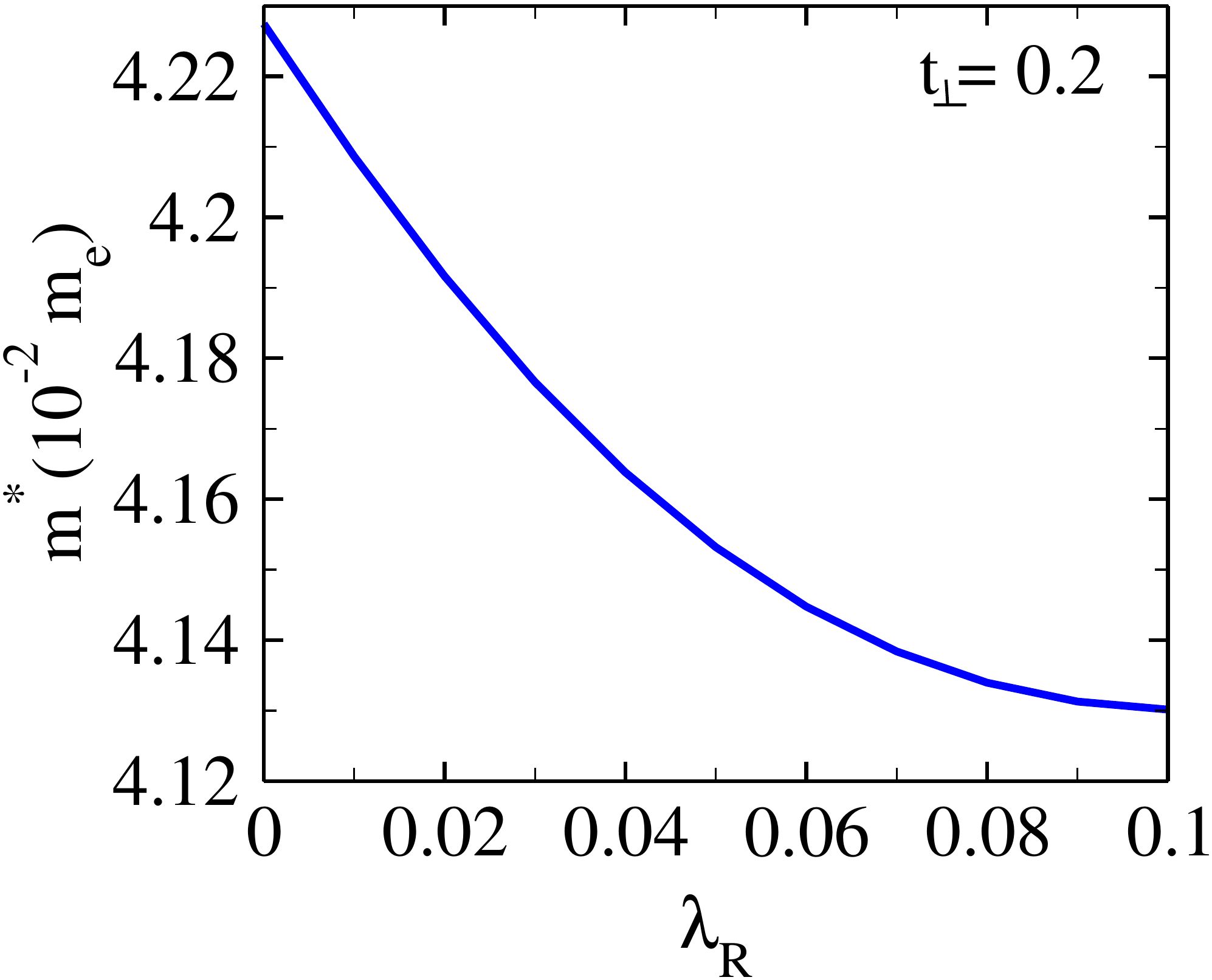}\label{fig:15}}
  \caption{(color online) Effective mass, $m^{*}$ as a function of intralayer Rashba spin orbit coupling parameter, $\lambda_R$ for zigzag bilayer graphene. Here, we set $t_\perp=0.2$, $t_2$ and $\lambda_R^{\perp}$ both are zero.}
\label{fig:15}
\end{figure}
In Fig.~\ref{fig:15} we have calculated the effective mass as a function of the intralayer Rashba SOC parameter, $\lambda_R$ for the lowest lying energy band (closest to the Fermi level $E=0$ and $N=10$) in the vicinity of a Dirac point. We have considered the inplane hopping, $t=2.7$ eV and the lattice constant, $a_0=2.46$  \AA. The energies are measured in units of $t$. We have not incorporated any intrinsic SOC and interlayer Rashba coupling here. The effective mass is seen to decrease with increase of the Rashba SOC. For $\lambda_R=0$, we have obtained some overestimated value of effective mass ($0.043 m_e$) which may be due to the artifact of the behaviour of the bandstructure of bilayer graphene. Consequently, the electrons will have larger group velocity for the corresponding band. As a result, the mobility of the electrons, which varies inversely with $m^{*}$ ($\mu \sim |m^{*}|^{-3/2} $)\cite{zhao}, increases.

\section{Conclusions}
\label{conclusion}
In this work, we derive analytical expressions for the edge
modes for a bilayer Kane-Mele model in presence of both SOCs. 
The analytic results show that the behavior of the edge states 
of a bilayer graphene is quite different than that of a monolayer 
graphene. For the band structure, the four bands correspond to 
four edge states which implies that there exists two edge modes 
per edge. An asymmetry in finite-size ribbon is also observed in presence of 
intrinsic SOC which otherwise is absent for a tight-binding model. With the inclusion of Rashba SOC, the band structure 
plots reveal that the QSH phase is destroyed in presence of an interlayer RSOC. Moreover, the charge conductance spectra show 
plateaus at $=4e^2/h$ for pristine bilayer graphene near the zero of the Fermi energy, while it decreases in presence of both intrinsic and Rashba SOCs. Studies on spin 
transport reveal that there are all non-zero components of spin 
polarization for a bilayer graphene which should be a positive input for spintronic applications. To make a connection with experiments we have computed the effective mass and have shown that it can be tuned by the inclusion of Rashba SOC.

\section{ACKNOWLEDGMENTS} 
SB thanks SERB, India for financial support under the grant F. No: EMR/ 2015/001039.

\end{document}